\documentclass[journal]{IEEEtran}
\usepackage{cite}
\usepackage{amsmath,amssymb}
\usepackage{graphicx}
\usepackage{xcolor}
\usepackage{multirow}
\usepackage[font=footnotesize]{caption}
\usepackage{stfloats}
\usepackage{todonotes}
\makeatletter
\def\endthebibliography{%
  \def\@noitemerr{\@latex@warning{Empty `thebibliography' environment}}%
  \endlist
}
\makeatother

\begin{document}
\bstctlcite{IEEE:BSTcontrol}

\title{A New k-Space Model for Non-Cartesian~Fourier~Imaging }

\author{Chin-Cheng Chan, \IEEEmembership{Student Member, IEEE} and Justin P. Haldar, \IEEEmembership{Senior Member, IEEE}
\thanks{This work was supported in part by NIH research grants U01-HL167613 and R01-MH116173, a USC Viterbi/Graduate School Fellowship, and computing resources from the USC Center for Advanced Research Computing.}
\thanks{C.-C. Chan and J. Haldar are with the Signal and Image Processing Institute, University of Southern California, Los Angeles, CA, USA (e-mail: chinchen@usc.edu; jhaldar@usc.edu).}}

\maketitle

\begin{abstract}
For the past several decades, it has been popular to reconstruct Fourier imaging data using model-based approaches that can  easily incorporate physical constraints and advanced regularization/machine learning priors.  The most common modeling approach is to represent the continuous image as a  linear combination of shifted ``voxel" basis functions. Although well-studied and widely-deployed, this voxel-based model is associated with  longstanding limitations, including high computational costs, slow convergence, and a propensity for artifacts.  In this work, we reexamine this model from a fresh perspective, identifying new issues that may have been previously overlooked (including undesirable approximation, wrap-around, and nullspace characteristics).  Our insights motivate us to propose a new model that is more resilient to  the limitations (old and new) of the previous approach. Specifically, the new model is based on a  Fourier-domain basis expansion rather than the standard image-domain voxel-based approach.  Illustrative results, which are presented in the context of non-Cartesian MRI reconstruction, demonstrate that the new model enables improved image quality (reduced artifacts) and/or reduced computational complexity (faster computations and improved convergence).
\end{abstract}


\section{Introduction}
\label{sec:intro}
In an ideal Fourier imaging experiment, data measurements are obtained by sampling the Fourier transform $F(\mathbf{k})$ of an unknown continuous image $f(\mathbf{x})$. In the noiseless case, this can be expressed as $d_m = F(\mathbf{k}_m)$ for $m=1,\ldots,M$, where $d_m\in\mathbb{C}$ is the $m$th measured data sample acquired at the Fourier (``k-space") sampling location $\mathbf{k}_m$, and
\begin{equation}
  \label{eq:forward}
  F(\mathbf{k}_m) \triangleq \iint_{-\infty}^{\infty} f(\mathbf{x}) e^{-i 2 \pi \mathbf{k}_m \cdot \mathbf{x}} d{\mathbf{x}}.
\end{equation}
Image reconstruction is the task of estimating $f(\mathbf{x})$ based on a noisy version of such data.   For simplicity, our subsequent descriptions will largely be written for the 1D case (using scalars $x$ and $k$ instead of the vector notation above), with straightforward generalizations to higher dimensions.

In recent decades, model-based image reconstruction methods \cite{fessler2010} have been popular  for their ability to easily combine data consistency constraints (involving an explicit model of data acquisition) with  prior information (e.g., physical constraints, advanced regularization, data-driven learned priors, etc.).  In this context, it has become common to represent $f(x)$ using a finite-dimensional linear basis expansion \cite{gordon1974,censor1983} of the form:
\begin{equation}
  \label{eq:img-model-general}
  f(x) = \sum_{n=1}^{N} b_n \eta_n(x),
\end{equation}
with coefficients $b_n$ and basis functions $\eta_n(x)$.  This representation allows Eq.~\eqref{eq:forward} to be rewritten in matrix-vector form as
\begin{equation}
\mathbf{d} = \mathbf{A} \mathbf{b},
\end{equation}
where $d_m$ and $b_n$ are respectively collected into the vectors $\mathbf{d} \in \mathbb{C}^M$ and $\mathbf{b}\in\mathbb{C}^N$, and the matrix $\mathbf{A}\in \mathbb{C}^{M \times N}$ has entries
\begin{equation}
[\mathbf{A}]_{mn} = \int_{-\infty}^\infty \eta_n(x) e^{-i2\pi k_m\cdot x} dx.
\end{equation}
Under white Gaussian noise assumptions, this naturally leads to  model-based reconstruction formulations of the form
\begin{equation}
\hat{\mathbf{b}} = \arg\min_{\mathbf{b} \in \mathbb{C}^N} \|\mathbf{A}\mathbf{b} - \mathbf{d}\|_2^2 + R_x(\mathbf{b}),
\label{eq:rec}
\end{equation}
where the penalty $R_x(\cdot)$ can be chosen to impose  prior information, and the reconstructed continuous image  is obtained by substituting $\hat{\mathbf{b}}$ into Eq.~\eqref{eq:img-model-general}.

While many image models can be used with Eq.~\eqref{eq:img-model-general} (including bespoke models with bases derived from subject-specific \cite{hu1988, liang1994, hess1999, tsao2001,tsao2003,liang2007}  or population-based \cite{cao1993} prior information), the   most-popular current approach is a generic voxel-based model, representing the image as a linear combination of uniform shifts of a ``voxel function'' $\varphi(x)$ with spacing $\Delta x$:
\begin{equation}
\label{eq:img-model}
f_v(x) = \sum_{n=-N/2}^{N/2-1} b_n \varphi(x - n \Delta x),
\end{equation}
where we assume $N$ is even for simplicity.\footnote{This 1D model is easily generalized to higher dimensions using tensor products of 1D functions.  For example, the common 2D version of this model is $f_v(x,y) = \sum_{n=1}^N b_n \varphi_x(x - p_n \Delta x)\varphi_y(y - q_n \Delta y)$, with $(p_n,q_n)\in\mathbb{Z}^2$.} The remainder of this paper will refer to Eq.~\eqref{eq:img-model} as the ``voxel-based model."
Common choices of $\varphi(x)$ include Dirac delta functions, sinc functions, and rectangle functions or other B-splines \cite{unser1999}.

Although Eq.~\eqref{eq:img-model}  is widely used,  it suffers from well-known limitations  when the k-space samples are non-Cartesian (i.e., the samples do not fall on a uniform lattice) \cite{pruessmann2001, fessler2003}. Specifically, it can be computationally expensive to evaluate the forward/adjoint operators (i.e., multiplying vectors by $\mathbf{A}$ or $\mathbf{A}^H$),  iterative image reconstruction methods can converge slowly, and it is common (though not always well-understood) to observe structured artifacts near the edges of the image.

The research community has invested decades of effort to minimize these  limitations. For instance, the complexity of multiplying vectors by $\mathbf{A}$ or $\mathbf{A}^H$ can be reduced by using approximations (e.g., gridding  \cite{pruessmann2001,osullivan1985, jackson1991, fessler2003, beatty2005,fessler2007}), by exploiting the convolutional structure of $\mathbf{A}^H\mathbf{A}$ \cite{wajer2001, fessler2005}, and/or by leveraging specialized computation hardware \cite{stone2008a,gai2013, baron2018}.   Similarly, convergence speed can be improved by using density compensation heuristics (at the expense of SNR-optimality)~\cite{pruessmann2001}, preconditioning \cite{trzasko2014,ong2020,iyer2024}, and/or better optimization algorithms \cite{weller2014, hong2024a}. Even with these techniques, the complexity of non-Cartesian reconstruction is still burdensome in many applications.  Moreover, while edge artifacts can  be mitigated by using stronger regularization and/or a larger field-of-view (FOV), this often comes at the expense of reduced image sharpness or increased computation.

Unlike  previous efforts that largely embrace the assumptions of Eq.~\eqref{eq:img-model}, our work examines this model skeptically, looking for  potential flaws and considering whether alternative models might exist that are equally general but have more favorable practical characteristics.  Our analysis leads to several realizations, including that Eq.~\eqref{eq:img-model} implicitly requires that the edge of k-space must wrap-around, and that the model has limited capacity to accurately represent the signal from some parts of the FOV.  We also observe that this model can be susceptible to producing unrealistic structured artifacts  in k-space, which may not have  been widely recognized in earlier work nor been attributed  to the use of a specific image model.

These insights motivate us to propose a general new model for Fourier imaging data, which adopts a linear basis expansion with basis functions that are localized in k-space.  Locality in k-space is desirable as it not only avoids the unrealistic k-space wrap-around of Eq.~\eqref{eq:img-model} and reduces sensitivity to  certain artifacts, but it also enables the use of computationally-efficient sparse matrix representations which have also proven beneficial in other settings \cite{rosenfeld1998, rosenfeld2002,yeh2005,liu2007,samsonov2008,kashyap2011}.  In addition, the new representation alters the distribution of subspace energy within the forward model, with potential benefits for the convergence of iterative algorithms.  Results shown later suggest that this approach enables improved modeling accuracy, reduced artifact sensitivity, and faster reconstruction.  Our proposed model is also fully compatible with modern iterative regularization methods and/or unrolled neural networks.   Highly-abbreviated preliminary accounts of portions of this work have been presented in recent conferences \cite{chan2025,chan2025a}.

This paper is organized as follows. Section~\ref{sec:limit-img-model} presents our analysis of the voxel-based model, focused on the new limitations we have identified. Section~\ref{sec:proposed_k_model} then introduces  our proposed Fourier-domain model, presents its theoretical characteristics, and discusses its commonalities with existing approaches.   The two models are compared empirically in the context of non-Cartesian MRI data in Sec.~\ref{sec:emp-recon-performance}. Discussion and conclusions are then presented in Sec.~\ref{sec:conclusion}.

\section{Analysis of the Image-Domain Voxel Model}
\label{sec:limit-img-model}

In the following subsections, we perform analyses that reveal several potentially undesirable features of  Eq.~\eqref{eq:img-model}.

\subsection{Eq.~\eqref{eq:img-model} and Fourier Wrap-Around}
\label{sec:peridicity}
Our first insight comes from a Fourier-domain perspective on Eq.~\eqref{eq:img-model}.  Specifically, it is straightforward  to show that the Fourier transform of Eq.~\eqref{eq:img-model} can  be expressed as
\begin{equation}
\label{eq:img-model-kbn}
\begin{split}
F_v(k) &\triangleq \int_{-\infty}^\infty f_v(x) e^{-i2\pi k\cdot x} dx = \Phi(k) \sum_{n=-N/2}^{N/2-1}  b_n  e^{-i2\pi n \Delta x k}\\
&= \Phi(k) \sum_{n=-N/2}^{N/2-1}  \beta_n  \xi_N^{(\Delta x)}\left(k - \frac{n}{N \Delta x}\right),
\end{split}
\end{equation}
where $\Phi(k)$ is the Fourier transform of $\varphi(x)$, $\beta_q$ is the discrete Fourier transform (DFT) \cite{oppenheim1999} of $b_n$:
\begin{equation}
\beta_q \triangleq \sum_{n=-N/2}^{N/2-1} b_n e^{-i2\pi n q /N},
\end{equation}
and  $\xi_N^{(\alpha)}(k)$ is a Dirichlet kernel:
\begin{equation}
\xi_N^{(\alpha)}(k) \triangleq \frac{1}{N}  \frac{\sin(\pi N \alpha k) }{\sin(\pi  \alpha k)}e^{i \pi \alpha k}.
\end{equation}
As can be seen, the voxel-based model is equivalent to a k-space model that expresses the Fourier signal $F(k)$ as a linear combination of uniform shifts of the k-space function $\xi_N^{(\Delta x)}(k)$, modulated by the function $\Phi(k)$.  It is perhaps remarkable how much the structure of Eq.~\eqref{eq:img-model-kbn} resembles that of  Eq.~\eqref{eq:img-model}, just in opposite domains --  both involve uniform shifts of a given function, although functions $\varphi(x)$ and $ \xi_N^{(\Delta x)}(k)$ often have very different behavior.  

\begin{figure}[t]
    \centering
    	\includegraphics[width=0.48\textwidth]{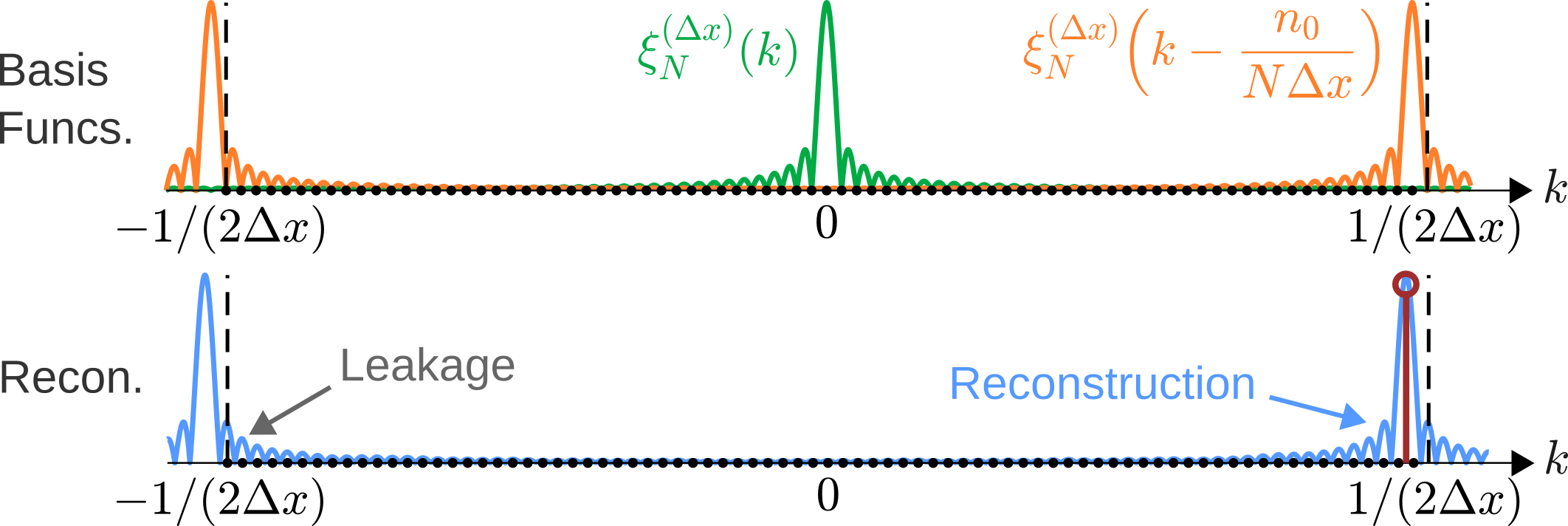}
    \caption{(top) Examples of k-space basis functions associated with Eq.~\eqref{eq:img-model}, corresponding to different shifts of $ \xi_N^{(\Delta x)}(k)$.   While the functions with small shifts near the center of k-space (green) have unremarkable characteristics, those with large shifts (orange) wrap around from one side of k-space to the other.  This can have undesirable consequences for image reconstruction, where (bottom) a simple minimum-norm least squares reconstruction (blue, obtained using $\hat{\mathbf{b}} = \mathbf{A}^\dagger \mathbf{d}$) of a single off-grid k-space sample (red) on one side of k-space results in signal leaking to the opposite side.  (Magnitude plots are shown, each curve has generalized linear phase).}
    \label{fig:periodicity}
\end{figure}

While Eq.~\eqref{eq:img-model-kbn} follows trivially from Eq.~\eqref{eq:img-model}, we are unaware of any prior  critiques of the structure of Eq.~\eqref{eq:img-model-kbn}  for model-based reconstruction or suggestions that it might be undesirable. From this perspective, it is important to note that Dirichlet kernels are sometimes called  ``periodic sinc functions" because of their periodicity characteristics -- in this case, $ \xi_N^{(\Delta x)}(k)$ is periodic with period $1/\Delta x$ for even $N$. This periodicity is relevant because the k-space basis functions in Eq.~\eqref{eq:img-model-kbn} are distributed across an interval that is also roughly $1/\Delta x$ in length (from $k\approx -1/(2\Delta x)$ to $k\approx +1/(2\Delta x)$). The result is that the basis functions corresponding to large $|n|$ must wrap-around from one side of k-space to the other, as illustrated in Fig.~\ref{fig:periodicity}.\footnote{For improved visualization, Fig.~\ref{fig:periodicity} uses a relatively small value of $N$ (i.e., $N=80$), although similar behavior is also observed with larger $N$.} This is potentially undesirable, as it forces unrealistic coupling between the two sides of k-space, which should normally be largely independent from one another in the absence of additional prior information.  Notably, this  can even cause information from one side of k-space to incorrectly leak to the other side of k-space during reconstruction, \footnote{Dirichlet structure and leakage of information from one side of k-space to the other can also occur for certain non-Cartesian Fourier reconstruction approaches that are not model-based, such as convolution gridding \cite{osullivan1985, fessler2007}.} as also illustrated in Fig.~\ref{fig:periodicity}.\footnote{In practice, this leakage effect may not have much practical impact with high-resolution data where there is little energy near the edge of k-space, although may have a more substantial impact in low-resolution scenarios.}  Of course, this problem could have been avoided with a different image model that did not induce  Dirichlet kernels in k-space.

\subsection{Eq.~\eqref{eq:img-model} and spatially-varying representation capacity}
\label{sec:capacity}
Our next insight comes from the observation that Eq.~\eqref{eq:img-model} has limited capacity to model all possible signals $F(k)$ that could potentially arise from the original continuous Fourier transform model (Eq.~\eqref{eq:forward}).  While this will be true of all finite-dimensional models that are used as approximate surrogates for infinite-dimensional continuous images, some models will have better approximation characteristics than others, and it can be helpful to gain insight into the nature  of the error.

\begin{figure}[t]
  \centering
  \includegraphics[width=0.48\textwidth]{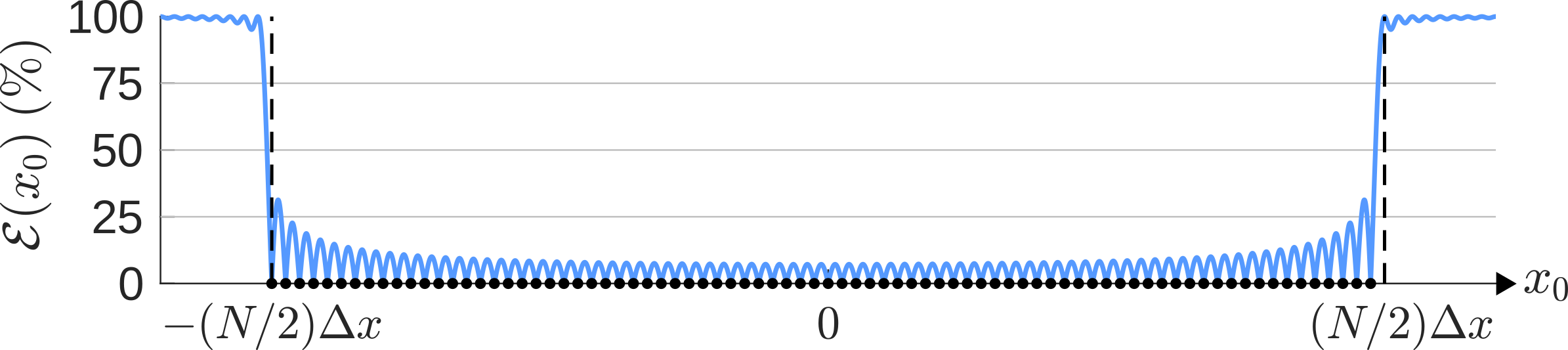}
  \caption{Plot of $\mathcal{E}(x_0)$, the best-case relative approximation error between the ideal signal arising from spatial location $x_0$ and the k-space signal model associated with Eq.~\eqref{eq:img-model}. Voxel locations are marked with black dots. }
  \label{fig:approx-error}
\end{figure}

In this work, we choose to examine Eq.~\eqref{eq:img-model}'s capacity to represent the signal arising from spatial locations both on and off-the voxel grid -- a type of model analysis that is well-established in the inverse problems literature\cite{chi2011}, and which is important because signal arising from off-grid locations will be present in all practical data, and the reconstruction process will attempt to fit it.  Due to the linearity properties of Eq.~\eqref{eq:forward} and because any $f(x)$ can be viewed as a linear superposition of Diracs (i.e., the sifting property of the Dirac delta function), we find it sufficient to consider spatial point sources $f_{x_0}(x)\triangleq \delta(x-x_0)$ for different $x_0$.  Through Eq.~\eqref{eq:forward}, such point sources give rise to complex sinusoids in k-space $F_{x_0}(k) \triangleq e^{-i2\pi k x_0}$.  Notably, while  the Dirichlet kernels in Eq.~\eqref{eq:img-model-kbn} must be periodic with period $1/\Delta x$, the sinusoidal signals $F_{x_0}(k)$  will have a mismatched period unless $x_0= n \Delta x$ for some integer $n\in \mathbb{Z}$.  Thus, we may expect good modeling accuracy for spatial locations $x_0$ that are perfectly aligned with the voxel grid from Eq.~\eqref{eq:img-model}, but should expect modeling errors for off-grid locations.  This type of problem could potentially be mitigated by using a different model without k-space periodicity characteristics.

We gain further insight by examining this behavior quantitatively.   For each spatial location $x_0$, define $\mathcal{E}(x_0)$ as the best-case relative  approximation error (in the $\mathcal{L}_2$-norm) between the ideal signal $F_{x_0}(k)$ and the k-space model $F_v(k)$ from Eq.~\eqref{eq:img-model-kbn} (corresponding to $f_v(x)$ from Eq.~\eqref{eq:img-model}), performing integration over the central period of  $F_v(k)$:
\begin{equation}
\label{eq:approx-error}
\hspace{-0.05in}\mathcal{E}(x_{0}) \triangleq  \min_{\mathbf{b}\in \mathbb{C}^{N}}\frac{\sqrt{\int_{-\frac{1}{2\Delta x}}^{\frac{1}{2\Delta x}} \left|  F_v(k) - F_{x_0}(k) \right|^2  dk }}{ \sqrt{\int_{-\frac{1}{2\Delta x}}^{\frac{1}{2\Delta x}} \left| F_{x_0}(k) \right|^2 dk}} \times 100\%.
\end{equation}
This optimization problem is easy to solve analytically using Hilbert space techniques \cite{luenberger1969,moon2000}, although the details are tedious and we omit them.

Figure~\ref{fig:approx-error} shows a plot of $ \mathcal{E}(x_{0})$ as a function of $x_0$ for a case where $\varphi(x)$ is chosen such that $\Phi(k) = 1$ within the central period of $F_v(k)$.\footnote{This choice is compatible with common choices of $\varphi(x)$, including Diracs ($\varphi(x) = \delta(x)$) and sinc functions ($\varphi(x) = \Delta x \sin(\pi x/ \Delta x)/(\pi x)$).} As expected, small approximation errors are observed when $x_0$ is close to the voxel grid (i.e., $x_0 \approx n \Delta x$ for $n\in \mathbb{Z}$), while errors increase as $x_0$ moves further from the grid locations. The approximation error grows especially large as $x_0$ moves away from the center of the FOV and goes towards $\pm \frac{N}{2}\Delta x$ (the edges of the voxel grid) and beyond.

These results suggest that Eq.~\eqref{eq:img-model} has limited capacity to accurately represent the signal from many parts of the FOV.  This is a potential concern, as real imaging data will generally include signal contributions from these regions. It should also be noted that these results represent the model's best possible approximation power (given oracle access to the true signal).  In a reconstruction scenario, the errors are potentially worse.

\subsection{Eq.~\eqref{eq:img-model} and structured k-Space artifacts}
\label{sec:cross_artifact}

Our last major insight about Eq.~\eqref{eq:img-model} comes from the empirical observation\footnote{We do not have a comprehensive theoretical understanding of these artifacts, which have multiple apparent origins, although believe that this is an interesting topic for future work.} that this model can be prone to producing  high-energy artifacts that appear along horizontal and/or vertical lines in k-space, often passing close to the k-space center. While image-domain artifacts are widely encountered in non-Cartesian imaging, we believe that structured k-space artifacts may not be well known, because the voxel-based model is not usually visualized in the Fourier domain. We came across them serendipitously during our investigations of Fourier-domain regularization  \cite{haldar2020} for non-Cartesian MRI reconstruction.

\begin{figure}[t]
  \centering
  \includegraphics[height=0.36\textwidth]{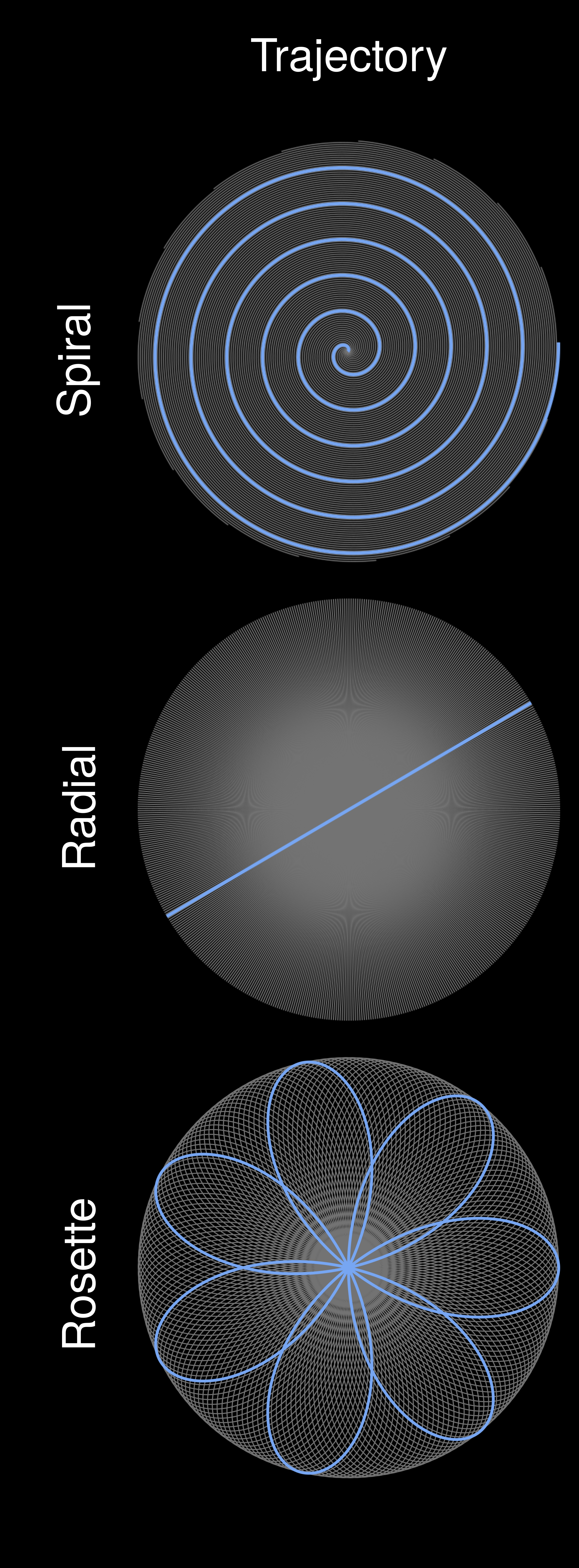}\includegraphics[height=0.36\textwidth]{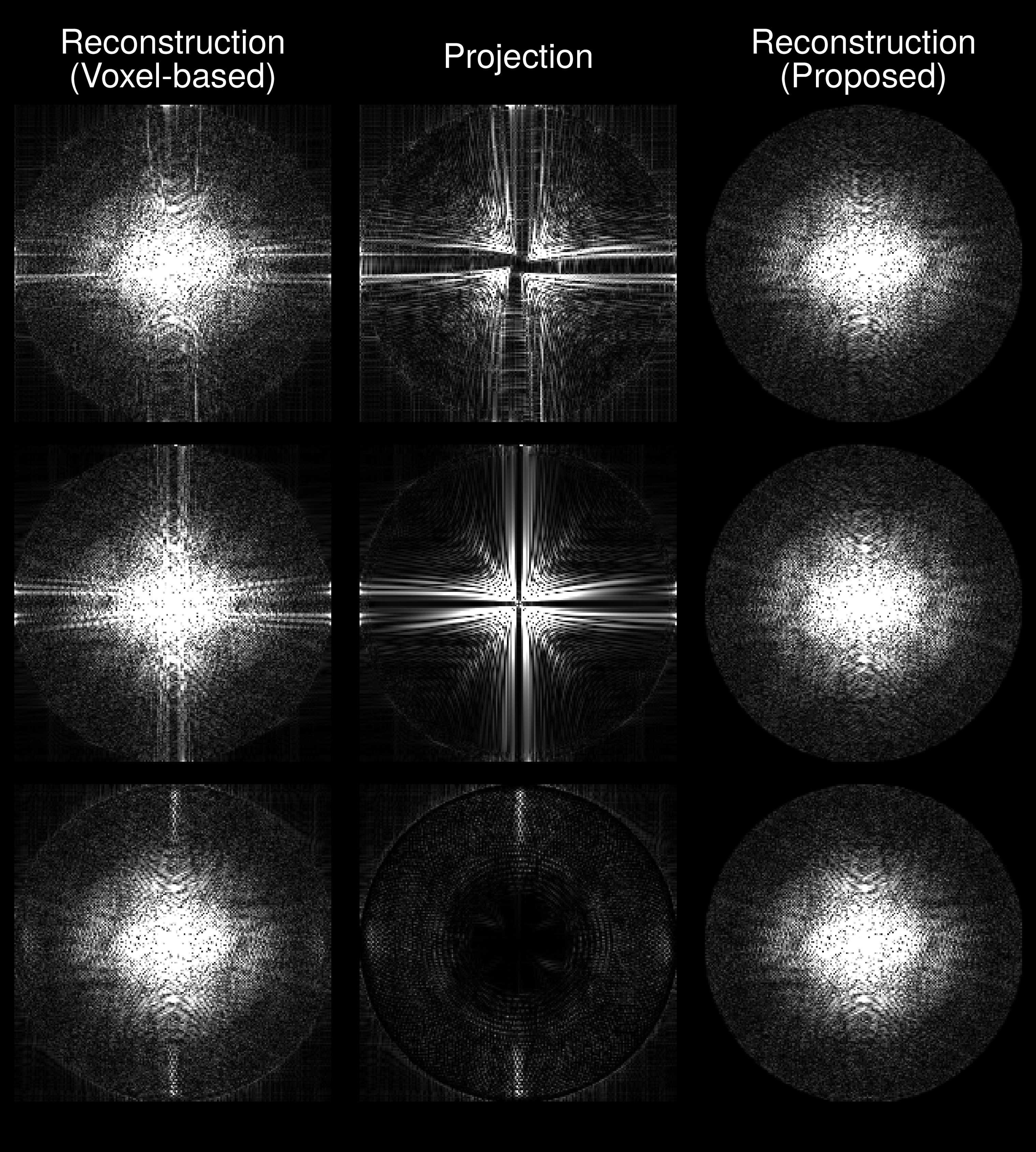}
  \caption{Structured k-space artifacts produced using the voxel-based model, for radial \cite{glover1992}, spiral \cite{meyer1992}, and rosette \cite{noll1997} trajectories. (left) k-space trajectories. (second column) Reconstructed k-space using Eq.~\eqref{eq:rec}. (third column) Projection of the reconstruction onto the near-nullspace of $\mathbf{A}$. For reference, we also show (right) reconstructed k-space using the proposed model.}
  \label{fig:cross_diff_traj}
\end{figure}

After our empirical observation of these artifacts with real data, we discovered that we could consistently induce them by simulating k-space data with signal contributions coming from outside the FOV (i.e., from entirely outside the spatial region spanned by the voxel grid), and then reconstructing that data using Eq.~\eqref{eq:rec} with Eq.~\eqref{eq:img-model} and weak regularization.\footnote{Notably, we have also observed these artifacts with real data in scenarios where we do not expect out-of-FOV signal, which leads us to believe that these artifacts can also arise in other scenarios.}  Figure~\ref{fig:cross_diff_traj} shows illustrative examples  across a range of different non-Cartesian k-space trajectories.\footnote{Figure~\ref{fig:cross_diff_traj} shows results using Dirac delta or sinc function basis functions, which produce equivalent results. We also observe similar artifacts when using rectangle basis functions.}  Here, Eq.~\eqref{eq:rec} was implemented with Tikhonov regularization (i.e., $R_x(\mathbf{x}) = \lambda \|\mathbf{x}\|_2^2$ with small regularization parameter $\lambda$), and $\mathbf{d}$ was comprised of simulated data from an analytic brain phantom \cite{guerquin2012} combined with analytic k-space \cite{pan1983}  from a low-intensity out-of-FOV ellipse. 

\begin{figure}[t]
\centering 
\includegraphics[width=0.48\textwidth]{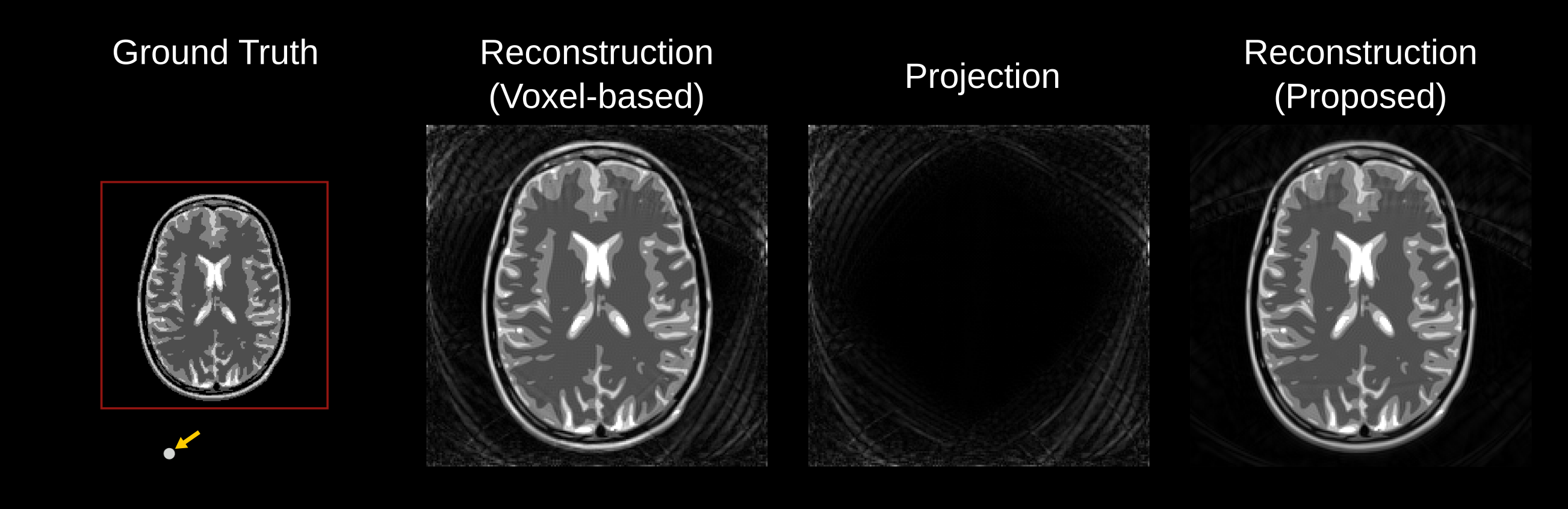}
\caption{The (left) image-domain version of the analytic phantom used in Fig.~\ref{fig:cross_diff_traj} (including an out-of-FOV ellipse, marked with a yellow arrow, with the nominal FOV marked in red), together with (right three columns) the image-domain versions of the spiral results from Fig.~\ref{fig:cross_diff_traj}. Note that we placed the ellipse far outside the FOV to avoid potential bias towards the proposed approach (which, as described later, uses a slightly ``extended" FOV), although similar artifacts also appear with the ellipse placed closer to the FOV edge. }
\label{fig:cross_img}
\end{figure} 

Our investigation suggests that, although the  structure and severity of these artifacts can vary substantially (depending on the trajectory, data, and reconstruction scheme), they consistently have most of their energy concentrated in the \emph{near-nullspace} of $\mathbf{A}$, i.e., the subspace associated with small singular values. This is illustrated in the third column of Fig.~\ref{fig:cross_diff_traj}, where we show the projection of the reconstruction onto the subspace associated with small singular values (i.e., singular values smaller than 5\% of the maximum singular value of $\mathbf{A}$ in most cases, using 10\% for the rosette). As can be seen, this projection captures most of the energy of these artifacts in all cases.  Notably, we also consistently observe that these artifacts are rapidly oscillating in k-space -- indeed, as illustrated in the image-domain results shown in Fig.~\ref{fig:cross_img}, when these k-space artifacts are transformed to the image domain, we consistently observe that their energy is concentrated near the edge of the FOV (i.e., the part of the FOV associated with fast k-space oscillations). This issue could potentially be mitigated using a model that has less capacity to represent rapid oscillations.

\section{A New k-Space Model}
\label{sec:proposed_k_model}

Motivated by our insights into Eq.~\eqref{eq:img-model}, we propose a new Fourier-domain model for the image that is based on a linear combination of uniform shifts of a k-space function $\Psi(k)$:
\begin{equation}
\label{eq:kspace-model}
F_\mathrm{k}(k) = \sum_{\ell=-L/2}^{L/2-1} c_\ell \Psi(k  - \ell\Delta k),\
\end{equation}
with coefficients $c_\ell$, model-order $L$, and basis spacing $\Delta k$.\footnote{As with Eq.~\eqref{eq:img-model}, the k-space model of Eq.~\eqref{eq:kspace-model} is presented in 1D for simplicity but is easily generalized to higher dimensions using tensor products.} The remainder of this paper will refer to Eq.~\eqref{eq:kspace-model} as the ``proposed k-space model." It is straightforward to show that $F_\mathrm{k}(k)$ has the equivalent image-domain representation:
\begin{equation}
\label{eq:k-model-image}
\begin{split}
\hspace{-.8em}f_\mathrm{k}(x) &\triangleq  \int_{-\infty}^\infty F_\mathrm{k}(k) e^{i2\pi k\cdot x} dk = \psi(x) \sum_{\ell=-L/2}^{L/2-1} c_\ell e^{i2\pi\ell \Delta k x}\\
&= L\psi(x) \sum_{\ell=-L/2}^{L/2-1} \gamma_\ell \xi_L^{(-\Delta k)}\left(x - \frac{\ell}{L\Delta k}\right),
\end{split}
\end{equation}
where $\gamma_q$ is the inverse DFT  \cite{oppenheim1999} of $c_\ell$. Notice also that, since the first line of Eq.~\eqref{eq:k-model-image} closely resembles a discrete-time Fourier transform (DTFT), it is easy to efficiently evaluate the image $f_\mathrm{k}(x)$ on an arbitrarily-dense grid of spatial locations $x$ by applying a zeropadded FFT to the coefficients $c_\ell$ \cite{oppenheim1999}, which can be helpful both for visualizing the reconstructed image and when using image-domain regularization penalties.

Notably, the proposed k-space model can be viewed as dual to the voxel-based model, simply interchanging the roles of $x$ and $k$.  Specifically, while the voxel-based model was associated with modulated Dirichlet kernels and wrap-around in k-space, the proposed k-space model is associated with modulated Dirichlet kernels and wrap-around in the image domain.  At first glance, it may not be obvious that this is an improvement, although we will demonstrate  that the proposed k-space model can have major practical advantages when $\Psi(k)$, $L$, and $\Delta k$ are chosen appropriately.

Before moving on, it will also be useful to observe that using this new model, Eq.~\eqref{eq:forward} can be simplified as:
\begin{equation}
\mathbf{d} = \mathbf{H}\mathbf{c},
\end{equation}
where $c_\ell$ is collected into $\mathbf{c} \in \mathbb{C}^L$, and  $\mathbf{H}\in\mathbb{C}^{M\times L}$ has entries:
\begin{equation}
[\mathbf{H}]_{m\ell} = \Psi(k_m  - \ell\Delta k).
\end{equation}
Similar to Eq.~\eqref{eq:rec}, under white Gaussian noise assumptions, this naturally leads to model-based reconstructions of the form:
\begin{equation}
\hat{\mathbf{c}} = \arg\min_{\mathbf{c} \in \mathbb{C}^L} \|\mathbf{H}\mathbf{c} - \mathbf{d}\|_2^2 + R_k(\mathbf{c}).
\label{eq:reck}
\end{equation}

\subsection{Selection of $\Psi(k)$}
\label{sec:sel-psi}
\begin{figure}[t]
    \centering
    	\includegraphics[width=0.48\textwidth]{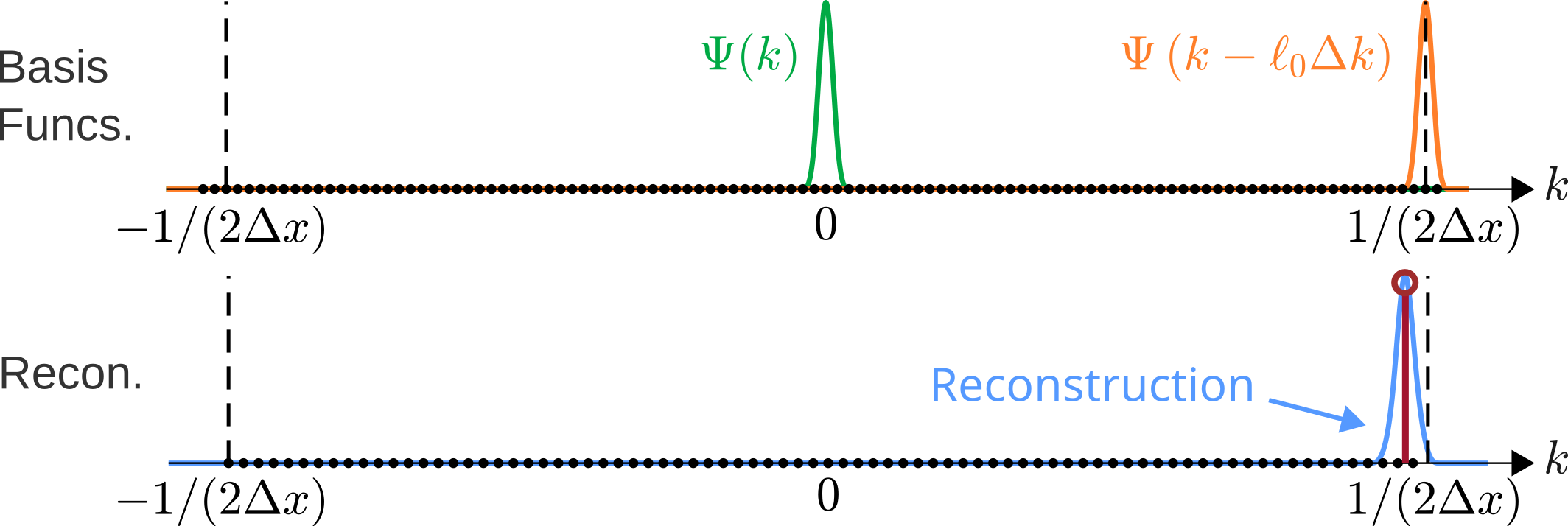}
    \caption{(top) Examples of different shifts of a compactly-supported nonperiodic k-space basis function $\Psi(k)$ (a third-degree B-spline in this case \cite{unser1999}) associated with Eq.~\eqref{eq:kspace-model}.  The wraparound seen in Fig.~\ref{fig:periodicity} is not present.  In addition, (bottom) we no longer observe signal leakage from one side of k-space to the other when performing minimum-norm least squares reconstruction (blue, obtained using $\hat{\mathbf{c}} = \mathbf{H}^\dagger \mathbf{d}$) of an off-grid k-space sample (red).}
    \label{fig:periodicity2}
\end{figure}

One key benefit of the proposed k-space model  is that we can choose  $\Psi(k)$ to be nonperiodic with compact/short support. This is unlike the functions $\xi_N^{(\Delta x)}(k)$ that are intrinsic to the k-space representation of Eq.~\eqref{eq:img-model}, which are  periodic and almost-everywhere nonzero in k-space, leading to  wrap-around and leakage as described in Sec.~\ref{sec:peridicity}. Figure~\ref{fig:periodicity2} shows that using different $\Psi(k)$ can avoid these issues.

In addition, choosing $\Psi(k)$ with short/compact support implies that any given k-space location $k_m$ will only depend on the small number of basis functions whose support includes $k_m$.  This enables the use of  sparse matrix representations for $\mathbf{H}$, which were not possible for Eq.~\eqref{eq:img-model} without approximation.

The function $\Psi(k)$ can also be chosen to avoid the type of structured k-space artifacts described in Sec.~\ref{sec:cross_artifact}.  As previously mentioned, these artifacts were always highly oscillatory.  Importantly, it was only possible for rapid oscillations to emerge because Eq.~\eqref{eq:img-model} has the capacity to produce rapid oscillations in k-space -- indeed, the Dirichlet kernels $\xi_N^{(\Delta x)}(k)$ associated with Eq.~\eqref{eq:img-model} are highly oscillatory themselves, and the artifacts are linear combinations of these kernels.  In principle, these  artifacts should be avoidable by choosing a function $\Psi(k)$ that is less oscillatory than $\xi_N^{(\Delta x)}(k)$.

There are many possible functions that are not periodic, have compact/short support, and are less oscillatory than $\xi_N^{(\Delta x)}(k)$, and therefore have the major features  desired for $\Psi(k)$ -- indeed, there are infinitely many such functions.  One natural choice would be to use B-spline basis functions \cite{unser1999}, which are well understood, are computationally simple to work with, have compact support, provide excellent approximation for families of smooth functions, and also  rise to and fall from a single peak without oscillation.

For simplicity and without loss of generality, the remainder of this paper will use the proposed k-space model   choosing $\Psi(k)$ to be a scaled B-spline basis function of degree $P$.  However, it is important to note that  Eq.~\eqref{eq:kspace-model} is fully compatible with other choices of $\Psi(k)$, and  the use of B-spline basis functions is merely illustrative and likely suboptimal.

For completeness, we provide a quick definition of B-spline basis functions below, referring readers to \cite{unser1999} for more detail.  Define the B-spline of degree 0 as the rectangle function:
\begin{equation}
\zeta_0(k) \triangleq \left\{ \begin{array}{ll} 1, & |k| \leq 1/2 \\ 0, & \text{else}. \end{array} \right.
\end{equation}
The B-spline of degree $P$  is obtained as the  $(P+1)$-fold convolution of $\zeta_0(k)$ with itself:
\begin{equation}
\zeta_P(k) \triangleq \underbrace{\zeta_0(k) \ast \zeta_0(k) \ast \ldots \ast \zeta_0(k).}_{(P+1) \text{ B-splines of degree 0}}
\end{equation}
Our later examples scale $\zeta_P(k)$ by $\Delta k$ to maintain  their desirable interpolation properties when the functions are spaced by $\Delta k$, i.e.,  $\Psi(k) = \zeta_P(k/\Delta k)$.  Notably, the zero-degree case $\zeta_0(k/\Delta k)$ corresponds to nearest-neighbor interpolation, while the first-degree case $\zeta_1(k/\Delta k)$  is the familiar triangle function corresponding to linear interpolation, and the third-degree case corresponds to popular cubic interpolation (which asymptotically provides an interpolant of minimum curvature  \cite{unser1999}).  The choice of $P$ represents a balance between compact support (good for efficient computation) and smoothness -- as $P$ grows larger,  $\zeta_P(k/\Delta k)$ grows increasingly smooth while its support grows larger (i.e., $\zeta_P(k/\Delta k)$ is only nonzero when $|k| <  (P+1)\Delta k/2$). In the sequel, we default to using $P=3$ (depicted in Fig.~\ref{fig:periodicity2}) unless otherwise noted.

The choice $\Psi(k) = \zeta_P(k/\Delta k)$ is equivalent to using $\psi(x) = \Delta k(\sin(\pi x \Delta k)/(\pi x \Delta k))^{P+1}$ in the image-domain (Eq.~\eqref{eq:k-model-image}).  This $\psi(x)$ is not only computationally simple to evaluate, but it also decays rapidly for large values of $x$, which will damp the out-of-FOV contributions from the periodic Dirichlet kernels in Eq.~\eqref{eq:k-model-image} when $|\psi(x)|$ is small beyond the nominal FOV.

In addition, it has recently been observed  that the choice of $\psi(x)$ can be implicitly linked to imposing prior information about the expected energy distribution of the original image $f(x)$ \cite{haldar2024}.\footnote{Indeed, this observation provides a promising mechanism for improving the choice of $\Psi(k)$ above and beyond using simple B-splines!}  In this case, the use of a function $\psi(x)$ that decays as $|x|$ grows large can be linked to an implicit prior that $f(x)$ is expected to have more energy near the center of the FOV than at the outskirts, which can be a good assumption for some imaging applications.  Note that, as will be shown later (cf. Sec.~\ref{sec:subspace-energy}), this implicit emphasis on the center of the FOV may also offer certain practical advantages, even if the image does not have more energy near the center of the FOV.

In scenarios where the interesting part of the image is not centered with respect to the FOV, we can also use the shift property of the Fourier transform to ``center" the image.  Specifically, instead of choosing $F_\mathrm{k}(k)$ to approximate $F(k)$, we can instead choose $F_\mathrm{k}(k)$ to approximate $e^{-i 2\pi  k \cdot x_0 }  F(k)$, noting that the inverse transform of $e^{-i 2\pi  k \cdot x_0 }  F(k)$ corresponds to the shifted image $f(x-x_0)$, and $x_0$ can be chosen so that the shifted image is centered.  In practice, this can be equivalently implemented by replacing $\mathbf{d}$ in  Eq.~\eqref{eq:reck} with $\mathbf{W}\mathbf{d}$, where $\mathbf{W}\in\mathbb{C}^{M\times M}$ is a diagonal matrix with $m$th diagonal entry $[\mathbf{W}]_{mm}=e^{-i 2\pi k_m \cdot x_0}$. Since $\mathbf{c}$ now models the ``centered'' image $f(x - x_0)$, evaluating the original uncentered image $f(x)$ requires replacing Eq.~\eqref{eq:k-model-image} by
\begin{equation}
  \label{eq:k-model-image-center}
  f(x) = \psi(x+x_0) \sum_{\ell=-L/2}^{L/2-1} c_\ell  e^{i2\pi\ell \Delta k (x+x_0)}.
\end{equation}

\subsection{Selection of $L$ and $\Delta k$}
The previous subsection described how we could choose $\Psi(k)$ in Eq.~\eqref{eq:kspace-model} to avoid some of the issues with the voxel-based model (i.e., wrap-around and structured artifacts, as discussed in Secs.~\ref{sec:peridicity} and \ref{sec:cross_artifact}, respectively). Herein, we discuss the selection of $L$ and $\Delta k$, which will be relevant to the representation capacity issues of Sec.~\ref{sec:capacity}.

To simplify  comparisons, we will first assume that parameters are chosen such that the basis spacing $\Delta x$ for the voxel-based model is matched with the image-domain ``voxel" spacing  observed in Eq.~\eqref{eq:k-model-image} for the proposed k-space model, i.e., $\Delta x= 1/(L\Delta k)$.  With this choice, the basis spacing in k-space  $1/(N\Delta x)$ (Eq.~\eqref{eq:img-model-kbn}) for the voxel-based model can be written as $1/(N\Delta x) = (L/N)\Delta k$.  Notably, this implies that the proposed k-space model has all of its $L$ Fourier-domain basis function positions uniformly spaced in the interval  $k\in [-(L/2) \Delta k , (L/2)\Delta k  - \Delta k]$, while the voxel-based model has all of its $N$ Fourier-domain basis function positions uniformly spaced in approximately the same interval $k \in [-(L/2) \Delta k, (L/2)\Delta k  - (L/N)\Delta k]$.  Thus, if we choose $L \geq N$ (as we often do), then the proposed k-space model can be viewed as effectively sampling the same k-space interval more densely than the voxel-based model, and we introduce $\rho \triangleq L/N$ as the ``oversampling factor" of the proposed k-space model. Note that the choice of $\rho$ and $N$ completely determines the values of $L$ and $\Delta k$.  In the image domain,  k-space oversampling means that the proposed k-space model has spatial basis function locations spanning an extended FOV ($\rho$-times larger than the FOV of the voxel-based model).

\begin{figure}[t]
  \centering
  \includegraphics[width=0.48\textwidth]{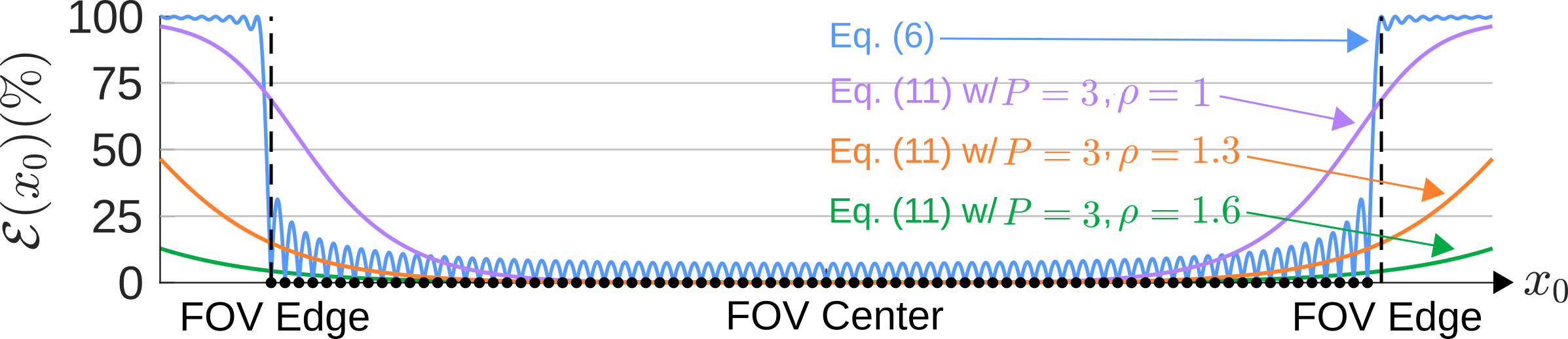}
  \caption{Plots of $\mathcal{E}(x_0)$ for different models associated with Eqs.~\eqref{eq:img-model} and \eqref{eq:kspace-model}. }
  \label{fig:approx-error2}
\end{figure}

To gain insight into the representation capacity of the proposed k-space model, Fig.~\ref{fig:approx-error2} shows plots of the best-case relative approximation error $\mathcal{E}(x_0)$ (i.e., Eq.~\eqref{eq:approx-error}, replacing $F_v(k)$ by $F_k(k)$) for different instances of the new model, with comparison against corresponding results (from Fig.~\ref{fig:approx-error}) for the voxel-based model.  As can be seen, representation capacity for the proposed k-space model varies smoothly with $x_0$, unlike the rapid oscillations observed with the voxel-based model.  In addition,  representation capacity improves as oversampling increases.  While using $\rho=1$ (no oversampling) arguably yields worse representation capacity than the image-domain voxel model (particularly near the outskirts), the representation capacity becomes substantially better as we start to oversample.  For example, the root-mean-squared value of $\mathcal{E}(x_0)$ (integrating over $x_0 \in [-(N/2) \Delta x, (N/2)\Delta x]$) was $4.1\%$ for the proposed model with $\rho = 1.3$, which is smaller than the value $11.2\%$ obtained for the voxel-based model.

\begin{figure}[t]
  \centering
  \includegraphics[width=0.48\textwidth]{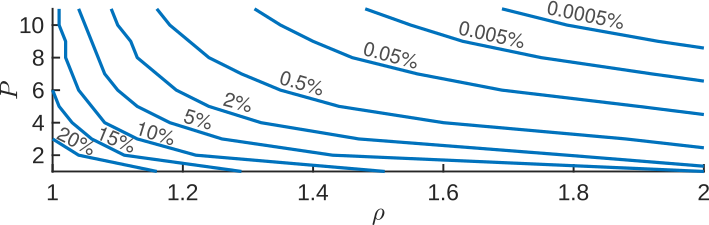}
  \caption{A contour plot showing the root-mean-squared value of $\mathcal{E}(x_0)$ as a function of $\rho$ and $P$ for the proposed k-space model.}
  \label{fig:choose-param}
\end{figure}

Figure~\ref{fig:approx-error2} was based on B-splines of degree $P=3$, although we should note that improvements in representation capacity can also be achieved using larger values of $P$, as depicted in Fig.~\ref{fig:choose-param}.   It should be noted that improving the representation capacity of the model using $P$ and $\rho$  is associated with  increased computational complexity in both cases (i.e., increasing $\rho$ increases the number of parameters to estimate and the number of columns of $\mathbf{H}$, while increasing $P$ increases the number of nonzeros appearing in each row of $\mathbf{H}$), so $P$ and $\rho$ should not be chosen unnecessarily large.

As already noted, we have often obtained good results when defaulting to $P=3$.  However,  selection of $\rho$ is more nuanced, depending on the characteristics of the data.  As can be seen from Fig.~\ref{fig:approx-error2}, signals from the center of the FOV can be accurately represented by the model even when $\rho$ is small.  As $\rho$ becomes larger, the size of this ``accurately modeled" spatial region increases.  This suggests that if $f(x)$ has limited support (with little energy close to the edge of the FOV), then it can be fine to choose smaller values of $\rho$, while using larger values of $\rho$ will be important to avoid representation errors for images with larger support.  This is also consistent with the interpretation of $\psi(x)$ as an implicit prior on the energy distribution (as discussed in the previous subsection), since $\psi(x)$ has its energy concentrated near the center of the FOV when $\rho$ is small, but this energy spreads over larger spatial region  as $\rho$ increases.  Our later results use $1 \leq \rho \leq 1.3$.

\begin{figure}[t]
  \centering
  \includegraphics[width=\linewidth]{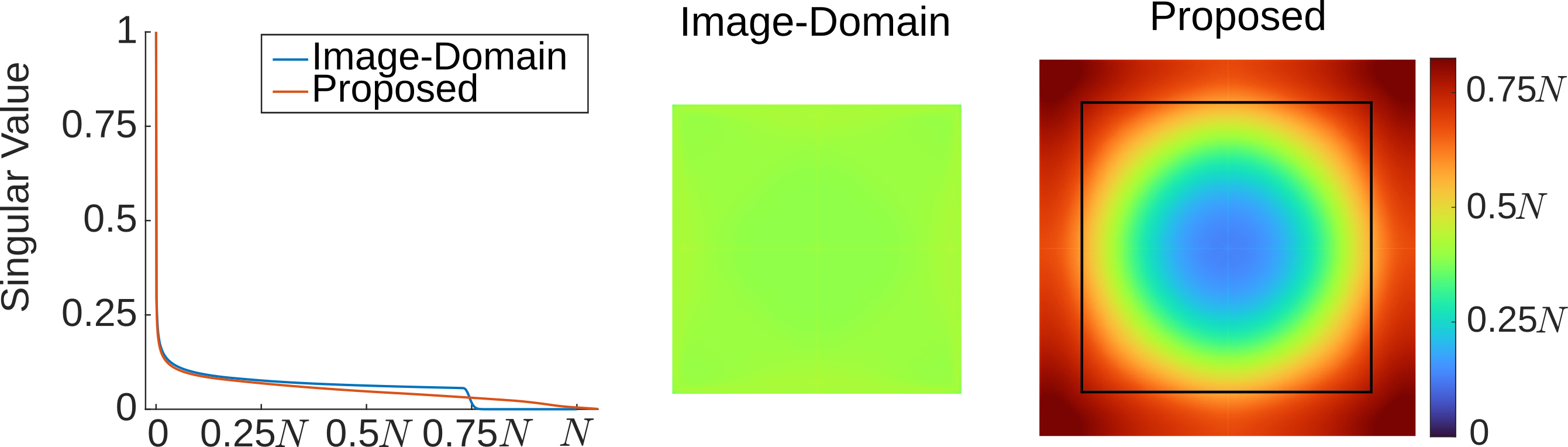}
  \caption{The $\mathbf{A}$ matrix associated with Eq.~\eqref{eq:img-model} and the $\mathbf{H}$ matrix associated with Eq.~\eqref{eq:kspace-model} have somewhat similar singular value profiles (left), although have very different subspace characteristics, as evident from calculating the mean singular value index corresponding to each spatial location (middle and right).   The k-space model used oversampling ($\rho=1.3$), resulting in a larger FOV -- the black square denotes the nominal (non-oversampled) FOV.}
  \label{fig:avg_sing_val}
\end{figure}

\subsection{Consequences for the distribution of subspace energy}
\label{sec:subspace-energy}
The choices of $\Psi(k)$, $L$, and $\Delta k$ described in the previous subsections were designed to avoid the specific issues we had identified in Sec.~\ref{sec:limit-img-model} for the voxel-based model. However, we also observe an unintended additional benefit, namely that these choices cause the matrix $\mathbf{H}$ to have its largest singular values more associated with spatial locations that are closer to the center of the FOV, while its smallest singular values are more associated with  the edges of the FOV.  This behavior is different than for the matrix $\mathbf{A}$ associated with Eq.~\eqref{eq:img-model}, which has a more uniform distribution of subspace energy. This is illustrated in Fig.~\ref{fig:avg_sing_val}, where we plot the  mean singular value index for each spatial location\footnote{ Let $\mathbf{A}$ have SVD $\mathbf{A} = \sum_i \sigma_i \mathbf{u}_i \mathbf{v}_i^H$, and note that the different entries of each vector $\mathbf{v}_i \in \mathbb{C}^N$ directly correspond to different spatial locations $n\Delta x$. Let $p_{n}(i) \triangleq  |[\mathbf{v}_{i}]_n|/(\sum_{i} |[\mathbf{v}_{i}]_n|)$ denote the normalized distribution describing the relative size of the contribution of each singular vector to the  $n$th spatial location. The mean singular value index for spatial location $n$ is obtained as $\mu_n \triangleq \sum_i i p_n(i)$. A similar approach is used for $\mathbf{H}$, after using a DFT to transform the matrix rows from k-space to spatial locations.}  for a radial trajectory (314 radial lines, with 200 samples per line) with 2D models ($N=200\times 200$, with $\rho=1.3$ for Eq.~\eqref{eq:kspace-model}).

This change to the subspace energy distribution likely occurs because of the way that $\psi(x)$ decays with increasing $|x|$. This subspace structure can be especially beneficial for avoiding ill-posedness and improving the convergence speed of iterative algorithms for experiments where the center of the FOV contains the most important information (which is common in many applications).  Specifically, it is widely observed (with strong theoretical foundations \cite{vogel2002}) that the solution components belonging to subspaces associated with the largest singular values are generally more robust to noise than components associated with smaller singular values.  In addition, when using iterative algorithms, these solution components often converge faster than others \cite{vogel2002}.

\subsection{Commonalities with Existing  Approaches}
Our proposed k-space model was obtained from a fresh perspective, starting from a different set of goals and objectives compared to previous work.  Nevertheless,  due to the underlying structure of the inverse problem, it may not be surprising that it has some commonalities with existing approaches.   We believe that our alternative way of thinking about the problem would be valuable even if it produced an image model that was ultimately equivalent to existing approaches that were derived from different perspectives. However, as explained below, our new model also has important distinctions. 

\subsubsection{Generalized Series} It is interesting to observe that our model can be viewed as a special case of both the classical Generalized Series model \cite{liang1994,hess1999} (especially its higher-order variant \cite{hernando2006}) and the later BLAST model \cite{tsao2001,tsao2003}.  While different realizations of these models exist, a common variation is that an image is represented by modulating a reference image  $\rho_{\mathrm{ref}}(x)$ with Fourier basis functions, i.e.,
\begin{equation}
f(x) = \rho_{\mathrm{ref}}(x) \sum_\ell c_\ell e^{i 2\pi \ell \Delta k x}.\label{eq:gs}
\end{equation}
Clearly, this representation matches that of Eq.~\eqref{eq:k-model-image} if we take $\rho_{\mathrm{ref}}(x)=\psi(x)$, such that our proposed model can thus be viewed as a special case of these earlier models.  However, it is important to note that these earlier models were proposed in a  different context than our approach.  Specifically, they were originally proposed in Cartesian imaging scenarios to enable high-quality image reconstruction from a limited number of Fourier measurements, with the reference image $\rho_{\mathrm{ref}}(x)$ used to impose constraints on the image reconstruction process to help compensate the lack of data. In particular, these approaches were designed so that the final reconstructed image would have an energy distribution and high-resolution spatial features that were similar to those present in $\rho_{\mathrm{ref}}(x)$.  In our case, $\psi(x)$ is chosen based on very different considerations that are specific to non-Cartesian imaging (as discussed in Secs.~\ref{sec:sel-psi}), although with potential benefits from the implicit energy distribution constraints (as discussed in Secs.~\ref{sec:subspace-energy}).

\subsubsection{Gridding}
Our proposed model also has some commonalities with gridding/non-uniform FFT methods \cite{osullivan1985, jackson1991, fessler2003, beatty2005,fessler2007}, which are frequently used to implement computationally-efficient approximations of the $\mathbf{A}$ matrix for the voxel-based model.  Specifically, these approaches approximate $\mathbf{A}$ as
\begin{equation}
\mathbf{A} \approx \mathbf{\Phi}\mathbf{G} \mathbf{F} \mathbf{Z} \mathbf{D},\label{eq:grid} 
\end{equation}
where $\mathbf{D} \in \mathbb{C}^{N\times N}$ is a diagonal matrix that applies spatially-varying ``deapodization" weights (derived from a function known as the gridding kernel) to the image prior to Fourier transformation; $\mathbf{Z} \in \mathbb{C}^{L \times N}$ is a zero-padding operation that expands the nominal FOV of the image prior to Fourier transformation; $\mathbf{F} \in \mathbb{C}^{L \times L}$ evaluates the Fourier transform of the image on an oversampled Cartesian grid in k-space; $\mathbf{G} \in \mathbb{C}^{M \times L}$ is a sparse matrix constructed from the gridding kernel to interpolate from the oversampled Cartesian grid points to the sampling locations $k_m$; and $\mathbf{\Phi} \in \mathbb{C}^{M\times M}$ is a diagonal matrix containing the $\Phi(k_m)$ values.  

Some elements of our proposed model bear resemblance to  steps of gridding.  Specifically, they both involve spatial weighting (i.e., $\psi(x)$ in Eq.~\eqref{eq:k-model-image} versus $\mathbf{D}$ in Eq.~\eqref{eq:grid}). However, $\psi(x)$ is a function that generally decays in magnitude away from the center of the FOV (akin to ``apodization") and is only used for post-reconstruction visualization of the image model (not in the numerical evaluation of $\mathbf{H}\mathbf{c}$). Conversely, $\mathbf{D}$ generally grows in magnitude away from the center of the FOV (``deapodization") and is directly used in image reconstruction (in the approximate numerical evaluation of $\mathbf{A}\mathbf{b}$), with no role in post-reconstruction visualization.  Moreover, the choices of $\psi(x)$ and $\mathbf{D}$ are made based on different considerations, with the gridding kernel (from which $\mathbf{D}$ is derived) chosen to ensure good approximation in Eq.~\eqref{eq:grid}, and with $\psi(x)$ chosen based on the distinct set of considerations from Sec.~\ref{sec:sel-psi}.

Both approaches also involve an oversampled Cartesian Fourier grid that is interpolated to non-Cartesian locations with a sparse matrix, although there are again important differences. In our proposed model, the image is defined over an FOV that extends beyond the nominal FOV, and reconstructed images usually  possess non-zero energy outside the nominal FOV.  In contrast, gridding relies on zero-padding to produce oversampling.  Importantly, zero-padding does not allow non-zero signal in the extended FOV, forcing the reconstructed image to be strictly contained within the nominal FOV.

From a bigger picture perspective, it is also important to note that these cosmetic similarities between certain low-level details of our proposed model and those of gridding do not imply that our model should behave similarly to a gridding-based implementation of the voxel-based model. Indeed, a key element of our contribution is that we are intentionally making a major shift in the function space of possible reconstructed images, moving away from the finite-dimensional subspace of continuous images used by the voxel-based model (which has fundamental limitations as identified in Sec.~\ref{sec:limit-img-model}) to a different subspace with better characteristics.  This shift results in major high-level changes in the way that reconstruction methods behave, in addition to any computational benefits it may afford.

\subsubsection{Sensitivity Modeling} During the peer review process for this paper, a reviewer suggested that the presence of $\psi(x)$  in Eq.~\eqref{eq:k-model-image} may have something in common with the use of sensitivity maps in multichannel MRI \cite{pruessmann1999, pruessmann2001}, particularly since sensitivity maps often taper to small or zero values away from the center of the image.  Although sensitivity maps are not typically used in single-channel scenarios (and the structure of the inverse problem changes dramatically when moving between single- to multi-channel scenarios), we will consider a single-channel perspective for simplicity.  

When using sensitivity maps, it is common to replace the $\mathbf{A}$ matrix with the matrix $\mathbf{E} = \mathbf{A}\mathbf{S}$  \cite{pruessmann1999, pruessmann2001}, where $\mathbf{S} \in \mathbb{C}^{N\times N}$ is a diagonal matrix containing the spatially-varying sensitivity map values.  To simplify our analysis, we will assume that $\mathbf{S}$ decays to small values away from the center of the FOV, but remains nonzero everywhere such that $\mathbf{S}$ is invertible.  In this scenario, it is notable that $\mathbf{E} \mathbf{b}$ is equivalent to $\mathbf{A}\widetilde{\mathbf{b}}$ under a change of variables $\mathbf{b} = \mathbf{S}^{-1} \widetilde{\mathbf{b}}$.  As a result, using $\mathbf{E}$ instead of $\mathbf{A}$ is no different than using the original voxel-based model, followed by rescaling the image by the inverse of the sensitivity map.   As a result, the use of sensitivity maps does not escape the previously-identified limitations of the voxel-based model (cf. Sec.~\ref{sec:limit-img-model}).  Moreover, since the sensitivity map decays to small values away from the center of the FOV, the rescaling of the image by the inverse of the sensitivity map is expected to amplify structure at the edges of the FOV, which is opposite from the behavior of the proposed model (in which the edges of the FOV tend to be damped). 

That said, an alternative approach to making the voxel-based model behave more like our proposed model using sensitivity map concepts might be to fuse Eq.~\eqref{eq:img-model} with the Generalized Series ideas from Eq.~\eqref{eq:gs}, i.e.:
\begin{equation}
f(x) = s(x) \sum_{n=-N/2}^{N/2-1} b_n \varphi(x - n \Delta x)
\end{equation}
where $s(x)$ is the sensitivity map/reference image, such that
\begin{equation}
F(k) = S(k) \ast \left[\Phi(k) \sum_{n=-N/2}^{N/2-1}  \beta_n  \xi_N^{(\Delta x)}\left(k - \frac{n}{N \Delta x}\right)\right],
\end{equation}
where $S(k)$ is the Fourier transform of $s(x)$ and $\ast$ denotes convolution.   This is somewhat akin to replacing $\mathbf{A}$ in Eq.~\eqref{eq:rec} with $\mathbf{E}=\mathbf{A}\mathbf{S}$ during reconstruction, and then using  $\mathbf{S} \hat{\mathbf{b}}$  for post-reconstruction visualization of the image \cite{hernando2006} (rather than just visualizing $\hat{\mathbf{b}}$ as is common for the conventional voxel-based model).  However, we are not aware of any prior literature that uses such an approach, and this approach will still suffer from some of the same limitations as Eq.~\eqref{eq:img-model} (e.g., k-space will still wrap-around) even if some of the other limitations are mitigated (e.g., the $S(k)$ convolution will damp the Dirichlet oscillations).

\section{Illustrative Comparisons}
\label{sec:emp-recon-performance}
In the following subsections, we compare the performance of the proposed k-space model against the voxel-based model in several representative 2D non-Cartesian MRI reconstruction scenarios (using 2D versions of the models). The first subsection focuses on conventional (single-channel) Fourier imaging as already described, while the second focuses on multichannel reconstruction (i.e., parallel imaging \cite{pruessmann2001, fessler2010}).

All reconstructions were implemented in MATLAB (without GPU acceleration) on a system with an 8-core Intel i7-9700 CPU and 64 GB RAM.  In all cases, we choose the voxel function $\varphi(\mathbf{x})$ for Eq.~\eqref{eq:img-model}  such that $\Phi(\mathbf{k}_m) = 1$ for all k-space sampling locations $\mathbf{k}_m$.  This choice does not uniquely specify the voxel function $\varphi(\mathbf{x})$ -- indeed, while this choice uniquely specifies $\mathbf{A}$ (and therefore also fully determines the optimal solution coefficients $\hat{\mathbf{b}}$),  $\Phi(\mathbf{k}_m) = 1$ is actually satisfied by infinitely many $\varphi(\mathbf{x})$ functions, including  choices such as Dirac delta functions and families of sinc and jinc functions. While the specific choice of $\varphi(\mathbf{x})$ does not influence  $\hat{\mathbf{b}}$, it will influence the visualization of  $\hat{f}_v(\mathbf{x})$ -- we visualize the image by displaying the $\hat{\mathbf{b}}$ coefficients directly, which can correspond to using a Dirac or sinc function for $\varphi(\mathbf{x})$.

In all cases, reconstructions were initially performed using a large number of iterations. We subsequently report computation time and convergence speed based on the time/iterations required to converge within a structural similarity (SSIM) \cite{wang2004} of 0.95 with the final converged result.\footnote{Note that, like all scalar metrics, SSIM does not capture all important aspects of image quality \cite{chan2023}. However, our results or conclusions do not change in important ways if we used other metrics such as normalized root-mean-squared error instead. }

\subsection{Standard (single-channel) Fourier reconstruction}
\label{sec:sing-chan-recon}
We first compare the behavior of Eq.~\eqref{eq:img-model} and Eq.~\eqref{eq:kspace-model} using single-channel reconstructions of two different $\sim$Nyquist-sampled non-Cartesian MRI datasets.  The first dataset was previously used in Ref.~\cite{stone2008a}, and corresponds to brain data acquired with a 3D stack-of-spirals trajectory (17 interleaves per slice with 3030 samples per readout, as shown in Fig.~\ref{fig:cross_diff_traj}) using a single-channel head coil on a 3T MRI scanner. The fully-sampled third dimension was reconstructed using the FFT, allowing independent 2D spiral reconstruction of each slice.  Reconstruction of this data is performed for a 256mm$\times$256mm FOV on an $N$=256$\times$256 grid for the image-domain voxel model (with $\rho=1.3$ for the proposed k-space model). The second dataset is publicly available \cite{lim2021}, and corresponds to vocal tract data acquired with a 2D spiral trajectory (13 interleaves with 630 samples per readout) using an 8-channel array coil on a 1.5T MRI scanner.  Reconstruction of the data from one representative coil was performed for a 200mm$\times$200mm FOV on an $N$=84$\times$84 grid for the image-domain voxel model (using $\rho=1.3$ for Eq.~\eqref{eq:kspace-model}).

\begin{figure*}[t]
  \centering
  \includegraphics[width=0.80\linewidth]{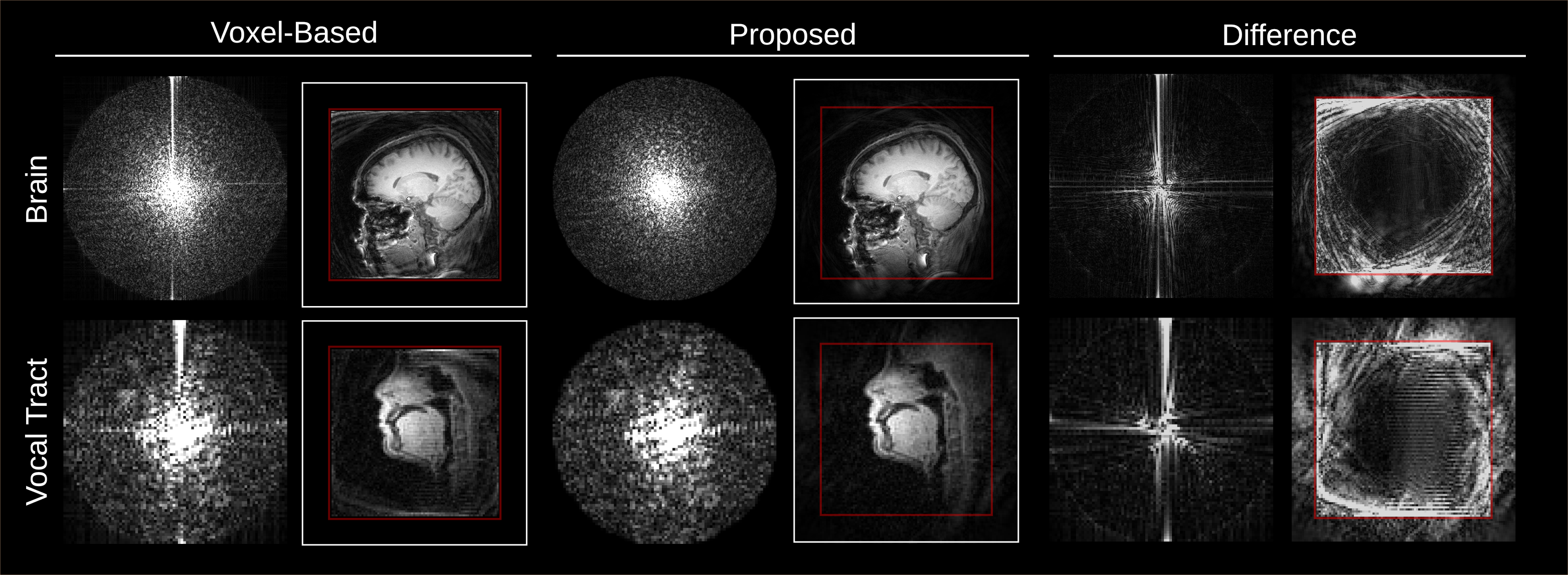}
  \caption{Reconstruction results from the voxel-based model and the proposed model, along with a visualization of the difference between the results obtained with the two models. Due to oversampling, the FOV for the proposed model is $1.3\times$ larger than for the voxel-based model. For easier comparison, we mark the nominal FOV and extended FOVs with red and white squares, respectively. Note that the intensity of the image-domain difference  has been scaled by 6$\times$ relative to the other visualizations of the image domain so that its features are easier to see.}
  \label{fig:spline_cross}
\end{figure*}

\begin{figure*}[t]
\centering
\includegraphics[width=0.8\linewidth]{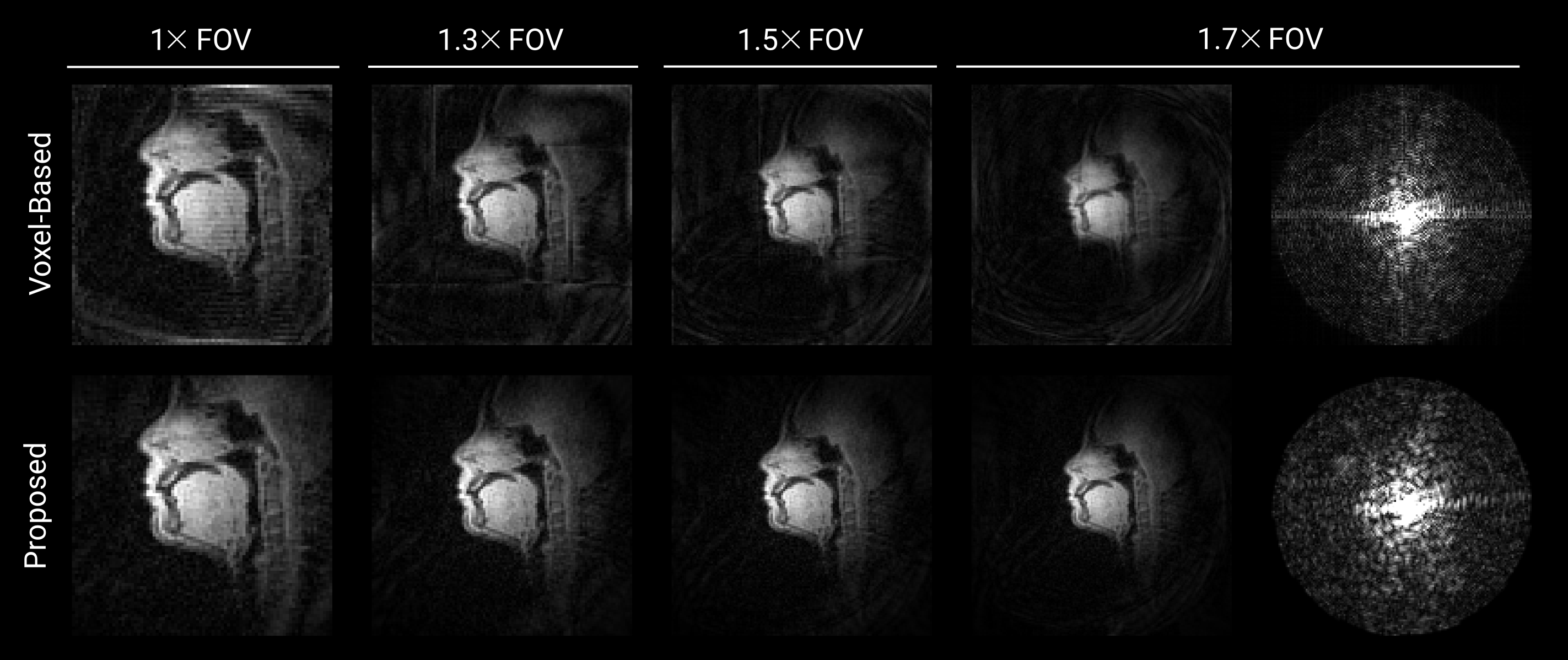}
\caption{Vocal tract reconstruction results obtained using different FOV sizes (cf. Fig.~\ref{fig:spline_cross}).}
\label{fig:vocal_track_recon}
\end{figure*}

Since data was $\sim$Nyquist sampled, reconstructions were performed using Eqs.~\eqref{eq:rec} and \eqref{eq:reck} with simple Tikhonov regularization for both models (i.e., $R_x(\mathbf{b}) = \lambda_x \|\mathbf{b}\|_2^2$ for Eq.~\eqref{eq:rec} and $R_k(\mathbf{c}) = \lambda_k \|\mathbf{c}\|_2^2$ for Eq.~\eqref{eq:reck}). Regularization parameters $\lambda_x$ and $\lambda_k$ were set as large as possible to reduce artifacts at the edge of the FOV, under the constraint that the regularization should not introduce perceptible blurring of anatomical structure. Reconstructions for both models were performed using both the conjugate gradient (CG) algorithm (using the Toeplitz approach \cite{wajer2001, fessler2005} to efficiently compute multiplications with $\mathbf{A}^H\mathbf{A}$, using 2$\times$-oversampled gridding  \cite{pruessmann2001, fessler2003, osullivan1985, jackson1991} to efficiently compute multiplications with $\mathbf{A}^H$, and using sparse matrices to perform multiplications with $\mathbf{H}$ and $\mathbf{H}^H$) as well as the LSQR algorithm\footnote{Theoretically, CG and LSQR should produce the same set of iterations with infinite numerical precision, with CG often slightly more computationally efficient. However, LSQR is generally more numerically stable and often demonstrates better convergence for problems that are more ill-conditioned.} \cite{paige1982} (as before, using gridding for $\mathbf{A}$/$\mathbf{A}^H$ and sparse matrices for $\mathbf{H}$/$\mathbf{H}^H$).

Reconstructed images obtained with CG are shown in Fig.~\ref{fig:spline_cross} (the images produced by CG versus LSQR were visually indistinguishable, although LSQR reached better cost function values). As can be seen, for both datasets, the voxel-based model produces a reconstruction with artifacts in both the image domain (i.e., artifacts near the edge of the FOV in both cases, with  ``ringing''-type stripes extending towards the center of the FOV for the vocal tract case) and in k-space (i.e., horizontal/vertical line artifacts), while the proposed k-space model does not exhibit these features.  To provide further insights, Fig.~\ref{fig:spline_cross} also shows images (in both k-space and the image domain) of the difference between the reconstruction results from each model.  As can be seen, the difference appears to be dominated by structured line-like artifacts in k-space and does not appear to contain any meaningful anatomical structure when visualized in the image domain.

One common approach to minimizing the impact of edge-of-FOV artifacts for the voxel-based model is to use a larger FOV (moving the artifacts further away from image features of interest,  at the expense of computational complexity).  While effective, we have  empirically observed that the selection of FOV can be nontrivial, requiring a trade-off between different types of artifacts.  Figure~\ref{fig:vocal_track_recon} shows reconstruction results using a range of FOVs (which were achieved by increasing $N$ while holding $\Delta x$ and the k-space trajectory fixed). As can be seen, FOV-selection for the voxel-based model must be done very carefully to avoid edge artifacts when the FOV is too small and blurring artifacts when the FOV is too large.  Interestingly, when using the largest FOV (which produced a blurry image), we observed that the corresponding k-space representation became highly-oscillatory, with low k-space intensities in between the turns of the spiral trajectory, perhaps reflecting an inability to accurately interpolate between k-space samples that are too far apart.\footnote{ Note that the sampling density of our spiral was anisotropic, with fast sampling along the direction of the spiral readout but larger distances radially between consecutive turns.  We believe that interpolation failures occurred in this case because the expanded FOVs interacted poorly with the low radial sampling density.  Specifically, expanding the FOV results in k-space basis functions that are packed together more tightly with smaller effective k-space ``footprints" (see Eq.~\eqref{eq:img-model-kbn}). This makes it easier for some k-space basis functions to largely reside in the gap between the k-space samples, which can make it more difficult to estimate reasonable basis function coefficient values.  } The same issues are not observed for the proposed k-space model, which is much less sensitive to the choice of FOV and exhibits fewer artifacts than the voxel-based model for every tested FOV size.

\begin{table}[t]
  \caption{Computational complexity (single-channel, CG)}
  \resizebox{\columnwidth}{!}{%
    \begin{tabular}{|l|cc|cc|cc|}
      \hline
      \multirow{2}{*}{} & \multicolumn{2}{c|}{Brain}            & \multicolumn{2}{c|}{Vocal Tract}      & \multicolumn{2}{c|}{Large FOV}        \\ \cline{2-7}
                        & \multicolumn{1}{c|}{Voxel} & Proposed & \multicolumn{1}{c|}{Voxel} & Proposed & \multicolumn{1}{c|}{Voxel} & Proposed \\ \hline
      Total Time (s)    & \multicolumn{1}{c|}{0.166} & \bf 0.040    & \multicolumn{1}{c|}{0.009} & \bf 0.005    & \multicolumn{1}{c|}{0.012} & \bf 0.007    \\ \hline
      Iters. to Convergence      & \multicolumn{1}{c|}{33}    & \bf 8        & \multicolumn{1}{c|}{15}    & \bf 9        & \multicolumn{1}{c|}{\bf 5}     & \bf5        \\ \hline
      Time per Iter. (ms)   & \multicolumn{1}{c|}{\bf 5.0}   & \bf 5.0      & \multicolumn{1}{c|}{0.6}   & \bf 0.5      & \multicolumn{1}{c|}{2.4}   & \bf 1.4      \\ \hline
    \end{tabular}%
  }
  \label{tab:convg-speed-cg}
\end{table}

\begin{table}[t]
  \caption{Computational complexity (single-channel, LSQR)}
  \resizebox{\columnwidth}{!}{%
    \begin{tabular}{|l|cc|cc|cc|}
      \hline
      \multirow{2}{*}{} & \multicolumn{2}{c|}{Brain}            & \multicolumn{2}{c|}{Vocal Tract}      & \multicolumn{2}{c|}{Large FOV}        \\ \cline{2-7}
                        & \multicolumn{1}{c|}{Voxel} & Proposed & \multicolumn{1}{c|}{Voxel} & Proposed & \multicolumn{1}{c|}{Voxel} & Proposed \\ \hline
      Total Time (s)    & \multicolumn{1}{c|}{0.783} & \bf 0.057    & \multicolumn{1}{c|}{0.063} & \bf 0.012    & \multicolumn{1}{c|}{0.049} & \bf 0.011    \\ \hline
      Iters. to Convergence      & \multicolumn{1}{c|}{39}    & \bf 8        & \multicolumn{1}{c|}{16}    & \bf 9        & \multicolumn{1}{c|}{6}     & \bf 5        \\ \hline
      Time per Iter. (ms)   & \multicolumn{1}{c|}{19.8}  & \bf 7.1      & \multicolumn{1}{c|}{3.9}   & \bf 1.3      & \multicolumn{1}{c|}{8.1}  & \bf 2.2      \\ \hline
    \end{tabular}%
  }
  \label{tab:convg-speed-lsqr}
\end{table}

Computational complexity is reported in Tables~\ref{tab:convg-speed-cg} and \ref{tab:convg-speed-lsqr} for CG and LSQR, respectively.  Results show that reconstructions with the proposed k-space model were faster than with the voxel-based model (between 1.7$\times$-4.0$\times$ faster for CG and 4.5$\times$-13.7$\times$ faster for LSQR), generally needing both fewer iterations to converge and less computation time per iteration.\footnote{Note that the use of the Toeplitz approach in CG for the voxel-based model means that multiplications with $\mathbf{A}^H\mathbf{A}$ do not scale with the number of k-space samples $M$, while the complexity of multiplying with $\mathbf{A}$, $\mathbf{A}^H$, $\mathbf{H}$, or $\mathbf{H}^H$ scales with $M$ for both gridding and sparse matrices.  In this $\sim$Nyquist scenario, $M$ is relatively large, which is advantageous for the Toeplitz approach, although advantages diminish with sub-Nyquist sampling.}   Consistent with Sec.~\ref{sec:subspace-energy}, we also observed that the center of the FOV converged more rapidly than the edges for the proposed k-space model.  We omit  details due to space constraints, noting evidence of similar behavior in the next subsection.

\subsection{Multichannel Fourier reconstruction}
\label{sec:fast-recon}
In MRI, it is frequent that Fourier data is acquired  simultaneously from an array of receivers.  In this context, ideal noiseless data acquisition can be represented as  \cite{pruessmann1999,pruessmann2001,fessler2010}
\begin{equation}
d_m^{(q)} = \iint_{-\infty}^{\infty} f^{(q)}(\mathbf{x}) e^{-i 2 \pi \mathbf{k}_m \cdot \mathbf{x}} d{\mathbf{x}}
\end{equation}
for $m=1,\ldots,M$ and $q=1,\ldots,Q$, where $Q$ is the number of receivers, $d_m^{(q)}$ represents the $m$th k-space sample acquired by the $q$th receiver, and the image $f^{(q)}(\mathbf{x})$ associated with the $q$th receiver channel corresponds to the true underlying image of interest $f(\mathbf{x})$ modulated by the sensitivity profile $s^{(q)}(\mathbf{x})$ of the $q$th receiver, i.e., $f^{(q)}(\mathbf{x}) = s^{(q)}(\mathbf{x})f(\mathbf{x})$.  This type of data acquisition can be beneficial since the sensitivity profiles provide an additional spatial encoding mechanism that can be used to enable reconstruction from sub-Nyquist k-space data.

Different reconstruction approaches exist for multichannel data, and we will present illustrative examples of two approaches.  The first approach imposes multichannel constraints directly in k-space, based on the existence of linear prediction relationships between the k-space samples of each channel \cite{haldar2020,griswold2002,lustig2010,haldar2014}.   The second approach (called SENSE \cite{pruessmann1999,pruessmann2001}) formulates reconstruction from an image-domain perspective, assuming sensitivity maps  are known. 

In both cases, multichannel reconstruction performance was assessed using publicly-available cardiac data \cite{maier2021} acquired with a highly-undersampled radial trajectory (27 radial lines with 320 samples per line) using a 34-channel array coil on a 3T MRI scanner. Data is prewhitened and compressed to 16 virtual channels using the SVD.\footnote{The transformation matrix used to achieve coil whitening/compression is unique up to rotation by a unitary matrix.  The choice of rotation matrix does not change the information contained in the virtual channels, although it does affect the spatial distribution of each channel. For the proposed k-space model, we prefer images with more compact energy distributions (cf. Sec.~\ref{sec:subspace-energy}).  As such, we applied varimax  \cite{kaiser1958} to the whitening/compression matrix, which finds a ``sparse" rotation matrix such that each virtual channel is approximately the combination of a small number of the original channels. This helps promote compactly-supported virtual channels, since the original channels are themselves spatially localized because of the array geometry. } Reconstruction was performed for a 480mm$\times$480mm FOV on an $N$=300$\times$300 grid for the image-domain voxel model (using $\rho=1$ for Eq.~\eqref{eq:kspace-model}, given the small support of each coil image). When needed, sensitivity maps were determined with PISCO \cite{lobos2023}.

Due to the high-degree of undersampling, regularization penalties were chosen to impose stronger priors than in the previous subsection.  Namely, LORAKS regularization \cite{haldar2014} (imposing support, phase, and multichannel correlation constraints) was used for multichannel k-space reconstruction, while total-variation (TV) regularization (imposing the constraint that the reconstructed image should have sparse gradients) was used with SENSE-based reconstruction.

\subsubsection{LORAKS} Non-Cartesian LORAKS reconstruction with the voxel-based model is typically formulated as
\begin{equation}
  \hat{\mathbf{b}}_{\mathrm{tot}} = \arg\min_{\mathbf{b}_{\mathrm{tot}} \in \mathbb{C}^{NQ}} \sum_{q=1}^Q  \|\mathbf{A}\mathbf{b}_q - \mathbf{d}_q\|_2^2 + J\left (\mathbf{K}_v \mathbf{b}_{\mathrm{tot}} \right),
  \label{eq:lrk-i}
\end{equation}
where $\mathbf{b}_q\in \mathbb{C}^N$  collects the image coefficients for the $q$th channel for $q=1,\ldots,Q$ (and $\mathbf{b}_\mathrm{tot} \in \mathbb{C}^{NQ}$ is the concatenation thereof), the matrix $\mathbf{K}_v$ samples the k-space representation of each channel image via Eq.~\eqref{eq:img-model-kbn},  $J(\cdot)$ is a regularization penalty that encourages a structured matrix formed from the k-space samples to possess low-rank characteristics, and we assume that multichannel noise was prewhitened \cite{pruessmann2001}.  This is straightforward to adapt to the proposed k-space model as:
\begin{equation}
  \hat{\mathbf{c}}_{\mathrm{tot}} = \arg\min_{\mathbf{c}_{\mathrm{tot}} \in \mathbb{C}^{QL}} \sum_{q=1}^Q  \|\mathbf{H}\mathbf{c}_q - \mathbf{W}_q\mathbf{d}_q\|_2^2 + J \left (\mathbf{V}_\mathrm{tot}^{-1}\mathbf{K}_k \mathbf{c}_{\mathrm{tot}} \right),
  \label{eq:lrk-k}
\end{equation}
where $\mathbf{c}_{q} \in \mathbb{C}^L$ collects the coefficients for the $q$th channel for $q=1,\ldots,Q$ (and $\mathbf{c}_\mathrm{tot} \in \mathbb{C}^{LQ}$  is the concatenation thereof); the matrices $\mathbf{W}_q\in\mathbb{C}^{M\times M}$ with entries $[\mathbf{W}_q]_{mm} = e^{-i2\pi k_m \cdot x_0^{(q)}}$ enable centering the signal from each channel  (cf. Sec.~\ref{sec:sel-psi}) in a channel-dependent way;  $\mathbf{V}_\mathrm{tot} \in \mathbb{C}^{LQ \times LQ}$ is the block-diagonal matrix formed from diagonal matrices $\mathbf{V}_q\in\mathbb{C}^{L\times L}$ with entries $[\mathbf{V}_q]_{\ell\ell} = e^{-i2\pi \ell \Delta k \cdot x_0^{(q)}}$ corresponding to the centering operations for the oversampled grid;\footnote{We use $\mathbf{V}_\mathrm{tot}^{-1}$ in Eq.~\eqref{eq:lrk-k} to ensure the multi-channel k-space samples used for LORAKS correspond to aligned multi-channel images, even though different centering shifts may be used for each channel.} and the matrix $\mathbf{K}_k$ samples the k-space of each channel via Eq.~\eqref{eq:kspace-model}. In practice, we have found it more efficient to simply use an identity matrix in place of $\mathbf{K}_k$,\footnote{Note that this is  equivalent to imposing LORAKS constraints on the ``unweighted" images $f^{(q)}(x)/\psi(x)$, which is reasonable since the unweighted images should have similar support, phase, and multichannel correlation characteristics to the $f^{(q)}(\mathbf{x})$ images.} and do this in our implementation.  We tested two variations of this approach, one without centering (i.e., each centering shift $x_0^{(q)}$ is zero) and one choosing the center shifts $x_0^{(q)}$ such that the support of each coil image is independently centered within the FOV.

\begin{figure}[t]
  \centering
  \includegraphics[width=1\linewidth]{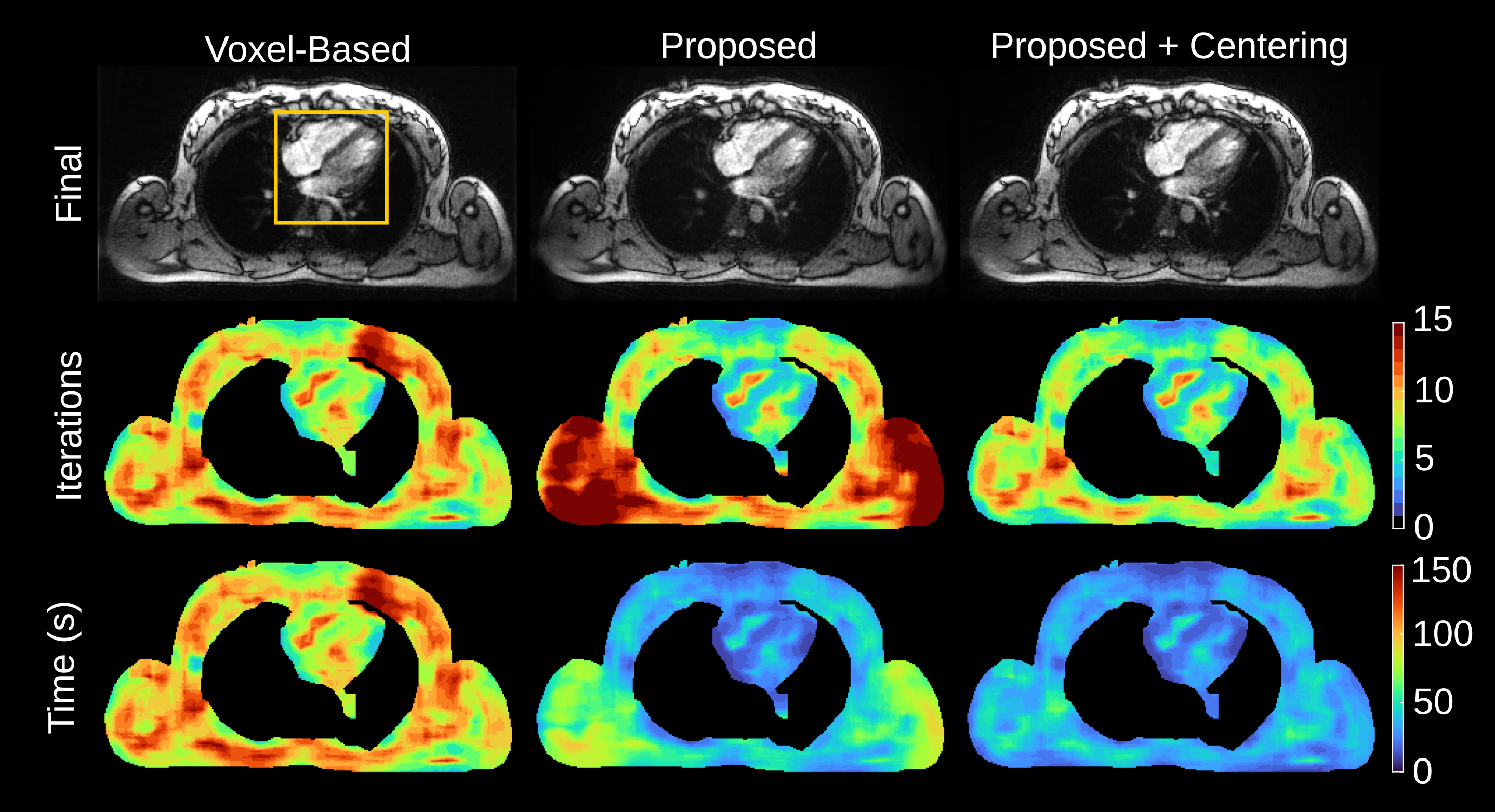}
  \caption{(top row) Final converged images obtained with LORAKS. The heart region used for speed calculations is marked with a yellow square.  (bottom two rows) Convergence characteristics as a function of spatial location.}
  \label{fig:loraks-final-convg}
\end{figure}

We used the P-LORAKS construction of the LORAKS $\mathbf{C}$-matrix with virtual conjugate coils \cite{Kim2019}, and the low-rank penalty function and algorithm from the original LORAKS paper \cite{haldar2014}. LORAKS parameters were selected to achieve the most qualitatively pleasing results for the voxel-based model, while these parameters were adjusted for the proposed k-space model to achieve a close match to the voxel-based results.

Images reconstructed with LORAKS are shown in the top row of Fig.~\ref{fig:loraks-final-convg}. As can be seen, all converged images are largely visually similar.\footnote{There is a minor difference observed on the left edge of the FOV, where a small edge artifact appears with the voxel-based model but not the proposed model -- but this is likely inconsequential.}  However, different regions of the FOV had different convergence characteristics.  The bottom two rows of Fig.~\ref{fig:loraks-final-convg} show the number of iterations and the amount of compute time required for different parts of the FOV to converge  (specifically, we show the computation required for the local 15$\times$15 window around each spatial location to converge within an SSIM value of 0.95 with the final converged result). These results show that the proposed k-space model (with or without centering) had faster convergence for regions near the center of the FOV while centering was beneficial for regions near the FOV edges. This is consistent with theoretical expectations (cf. Secs.~\ref{sec:sel-psi} and \ref{sec:subspace-energy}).  Computational complexity for the heart region \footnote{Note that it may be unnecessary to let the entire image converge if there is a subregion of interest that converges faster than other regions.} (cf. Fig.~\ref{fig:loraks-final-convg})) is reported in Table~\ref{tab:convg-speed-loraks}.  Results show that reconstruction with the proposed model (with centering) was  3.5$\times$ faster than with the voxel-based model (needing fewer iterations and less compute time per iteration).

\begin{table}[t]
  \caption{Computational complexity (multichannel, LORAKS)}
  \centering
  \begin{tabular}{|l|c|c|c|}
    \hline
    & Voxel & Proposed & Proposed+Centering \\ \hline
    Total time (s) & 85.4  & 28.0      & \bf 24.2                \\ \hline
    Iters. to Convergence   & 8    &  6       & \bf 5                 \\ \hline
    Time per Iter. (s) & 10.68   & \bf 4.67      & 4.83      \\ \hline
  \end{tabular}
  \label{tab:convg-speed-loraks}
\end{table}

\subsubsection{SENSE+TV} Non-Cartesian SENSE reconstruction with TV regularization is typically based on using Eq.~\eqref{eq:img-model} to model the underlying image $f(\mathbf{x})$, with Eq.~\eqref{eq:rec} replaced by:
\begin{equation}
\hat{\mathbf{b}} = \arg\min_{\mathbf{b} \in \mathbb{C}^N} \sum_{q=1}^Q \|\mathbf{A}\mathbf{S}_q\mathbf{b} - \mathbf{d}_q\|_2^2 + \lambda\|\mathbf{D}\mathbf{b}\|_1,
\label{eq:stv-reci}
\end{equation}
where $\mathbf{S}_q\in\mathbb{C}^{N\times N}$ is a diagonal matrix with diagonal entries $s^{(q)}(\mathbf{x}_n)$ to represent  sensitivity encoding, $\mathbf{D}$ is a spatial finite difference operator, and $\lambda$ is the regularization parameter.

There are multiple ways of adapting SENSE-type constraints for our proposed k-space model. One such approach is to note that under our proposed model, $f(x)$ can be rewritten as $f(x) = \psi(x) g(x)$ where $g(x)$ is defined as the Dirichlet series appearing on the right side of Eq.~\eqref{eq:k-model-image} (i.e., $g(x) \triangleq L \sum_{\ell=-L/2}^{L/2-1} \gamma_\ell \xi_L^{(-\Delta k)}\left(x - \frac{\ell}{L\Delta k}\right)$).   We then observe that the model coefficients $c_\ell$   have a one-to-one correspondence with the set of $L$ spatial samples of the Dirichlet series $g(\ell/(L \Delta k))$ for $\ell=-L/2,\ldots, L/2-1$  (i.e., $\mathbf{g}\in \mathbb{C}^L$), and we can choose to solve for $\mathbf{g}$ directly. This allows us to formulate SENSE+TV reconstruction as:
\begin{equation}
\hat{\mathbf{g}} = \arg\min_{\mathbf{g} \in \mathbb{C}^L} \sum_{q=1}^Q \| \mathbf{H}\mathbf{F}\mathbf{S}_q \mathbf{g} -\mathbf{d}_q\|_2^2 + \lambda\|\mathbf{D}\mathbf{Y}\mathbf{g}\|_1,
\label{eq:stv-reck}
\end{equation}
where $\mathbf{F} \in \mathbb{C}^{L\times L}$ is the length-$L$ DFT matrix (which can be implemented using conventional FFTs) and $\mathbf{Y}\in\mathbb{C}^{L\times L}$ is the diagonal matrix of $\psi(\ell/(L \Delta k))$ samples.

Optimization of Eqs.~\eqref{eq:stv-reci} and \eqref{eq:stv-reck} was performed using FISTA \cite{beck2009}. For the voxel-based model, $\lambda$  was tuned to achieve the most qualitatively pleasing results, while for the proposed k-space model, $\lambda$ was chosen to achieve a close match to the voxel-based results. 

\begin{figure}[t]
  \centering
  \includegraphics[width=0.72\linewidth]{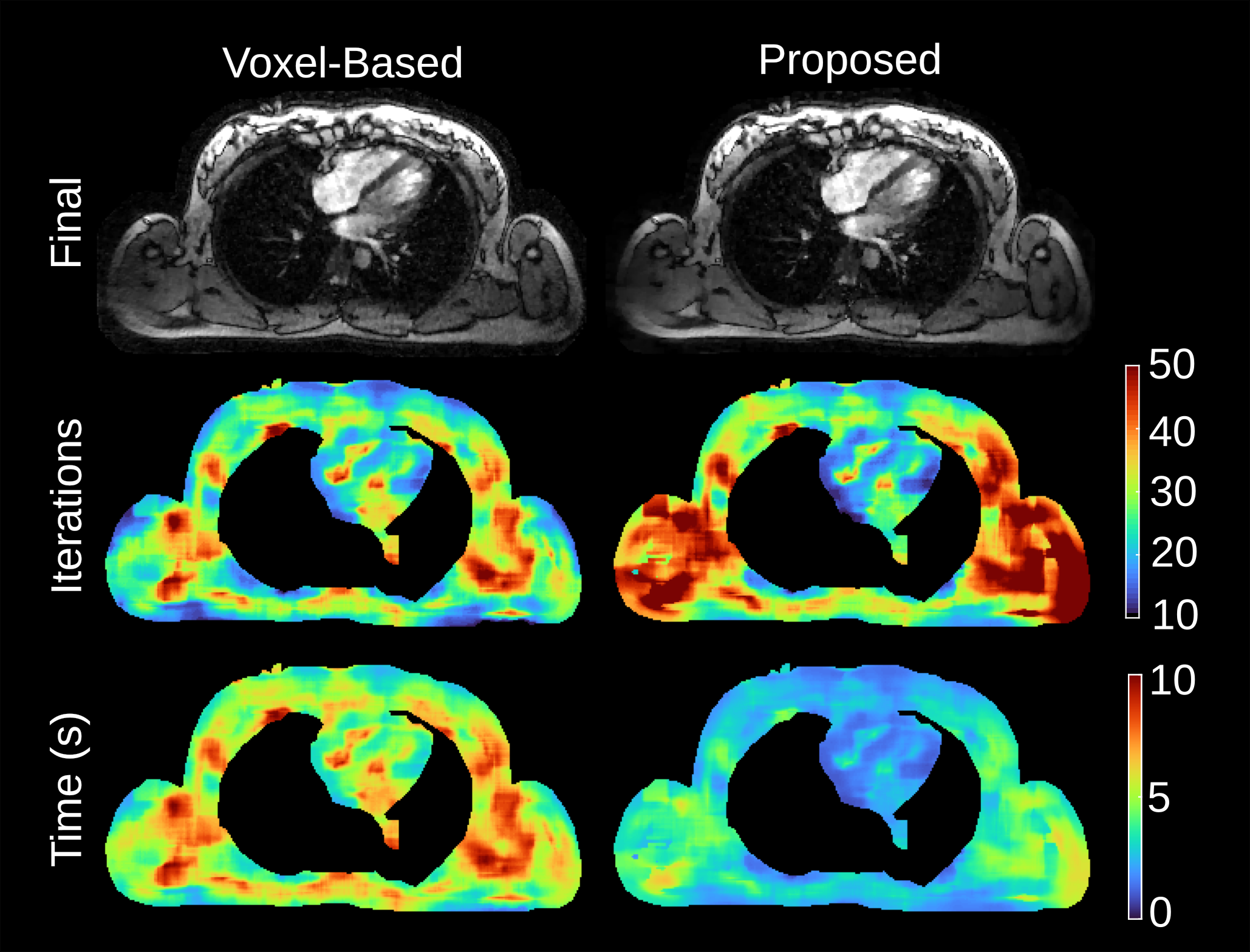}
  \caption{SENSE+TV results, displayed in the same format as Fig.~\ref{fig:loraks-final-convg}.}
  \label{fig:fista-final-convg}
\end{figure}

Reconstruction results and convergence characteristics are visualized in Fig.~\ref{fig:fista-final-convg}. As before, all converged images are visually similar, and we observe differences in convergence speed between the center and edge of the FOV for the proposed model.

Computational complexity for the heart region is reported in Table~\ref{tab:convg-speed-fista}.  In this case, we do not observe a very major difference in the number of iterations required to converge, although the proposed model still yields 2.9$\times$ faster reconstruction than the voxel-based model due to reduced per-iteration complexity.

\begin{table}[t]
  \caption{Computational complexity (multichannel, SENSE+TV)  }
  \centering
  \begin{tabular}{|l|c|c|c|}
    \hline
    & Voxel & Proposed  \\ \hline
    Total Time (s) & 5.2  & \bf 1.8                  \\ \hline
    Iters. to Convergence   & \bf 33    & \bf 28       \\ \hline
    Time per Iter. (s) & 0.16   & \bf 0.06          \\ \hline
  \end{tabular}
  \label{tab:convg-speed-fista}
\end{table}

\section{Discussion and Conclusions}
\label{sec:conclusion}

This work identified several new limitations of the widely-used voxel-based model (Eq.~\eqref{eq:img-model}) that do not appear to be well known. Our insights  allowed us to propose a new Fourier-domain model that avoids or mitigates the limitations of Eq.~\eqref{eq:img-model}.  Our theoretical analysis and empirical testing with MRI data suggest that this new model offers improved representation capacity, reduced vulnerability to artifacts, and improved computational efficiency. While we only reported a few examples due to space constraints, we have observed similar behavior across a wide range of different non-Cartesian Fourier reconstruction scenarios. Although we only examined the proposed model in 1D and 2D scenarios, we also expect the proposed model to be beneficial in  higher dimensions.  Overall, we anticipate that the proposed  model will be useful for a range of non-Cartesian MRI applications, and offer potential benefits to other Fourier imaging modalities more broadly.

This paper focused on higher-level image modeling concepts rather than lower-level implementation details.  Specifically, our results were based on a relatively-unoptimized MATLAB implementation of the proposed model, and we have left a detailed investigation of efficient software and hardware implementations to the future. Interestingly, we have anecdotally observed  that our current immature implementations of the proposed method can even have substantial speed advantages (though with memory disadvantages) when compared against the voxel-based model using  highly-refined CPU and GPU implementations of the non-uniform FFT \cite{barnett2019,shih2021}.  These results support the potential of the proposed approach, and suggest that the development of even more efficient implementations may be a promising endeavor.    

Notably, while our new model is compatible with a wide range of model-based reconstruction approaches, our evaluation intentionally focused on principle-driven regularizers that impose constraints such as sparsity and low-rank structure.  This is because such approaches only have a few degrees of freedom and are easier to investigate compared to data-driven regularization approaches based on machine learning techniques (such as deep unrolling or plug-and-play methods).  We anticipate that a thorough investigation of the impact of the new model on data-driven regularization would also be very interesting (with potentially major implications for both network training and network architecture), and hypothesize that major benefits might be obtained given the consistent advantages we observed in this work.  However, due to space constraints and due to the complexity and rigor that would be required to reach meaningful conclusions, we believe that such an investigation is beyond the scope of this paper and is best left to future work.

The ability to use sparse $\mathbf{H}$ is one of the factors contributing to the computational efficiency of the proposed model.  However, it should be noted that the computational complexity of $\mathbf{H}$ scales with the number of k-space samples $M$, and the proposed approach should be particularly  beneficial for sub-Nyquist acquisitions where $M$ is small.

Unlike the traditional voxel-based model, the proposed k-space model also strongly emphasizes the center of the FOV (if $\mathbf{H}$ is used in isolation). This  enables the center of the FOV to converge more rapidly when using iterative algorithms. The proposed model is also more robust to artifacts arising from poorly-chosen FOVs (e.g., enabling the use of smaller FOVs, thereby reducing computational complexity even further). 

The results presented in this paper were based on using cubic B-splines for $\Psi(k)$ -- one of infinitely many $\Psi(k)$ choices that fulfill the theoretical criteria we outlined  in Sec.~\ref{sec:proposed_k_model}.  An exploration of alternative basis functions, such as those discussed in Ref.~\cite{potts2001}, would be an interesting direction for future work.  Another interesting topic would be to explore the extent to which similar data-domain models may offer benefits over image-domain models in other scenarios (e.g., for sinograms in tomography, or for data measurements in Fourier-adjacent MRI scenarios involving field inhomogeneity, gradient nonlinearity, concomitant fields, etc.).

A sample software implementation of the proposed k-space model is available from: https://doi.org/10.5281/zenodo.20483042.

\bibliographystyle{IEEEtran}
\bibliography{./bibliography}

\end{document}